\documentclass{article}

\usepackage{arxiv}

\usepackage[utf8]{inputenc} 
\usepackage[T1]{fontenc}    
\usepackage{hyperref}       
\usepackage{url}            
\usepackage{booktabs}       
\usepackage{amsfonts}       
\usepackage{nicefrac}       
\usepackage{microtype}      
\usepackage{lipsum}
\usepackage{graphicx}
\usepackage{amsmath}
\usepackage{authblk}

\title{Light modulation in Silicon photonics by PZT actuated acoustic waves}

\author[1,2,3]{Irfan Ansari}
\author[2,3]{John P. George}
\author[1,2,3]{Gilles F. Feutmba}
\author[2,3]{Tessa Van de Veire}
\author[1,2]{Awanish Pandey}
\author[2,3]{Jeroen Beeckman}
\author[1,2]{Dries Van Thourhout}

\affil[1]{\textit{Photonics Research Group, INTEC Department, Ghent University-IMEC}}
\affil[2]{\textit{Centre for Nano and Bio-photonics, Ghent University}}
\affil[3]{\textit{Liquid Crystal \& Photonics Group, ELIS Department, Ghent University}}
\affil[*]{\textit{Corresponding author: dries.vanthourhout@UGent.be}}

\begin{document}
\maketitle


\begin{abstract}
Tailoring the interaction between light and sound has opened new possibilities in photonic integrated circuits (PICs) that ranges from achieving quantum control of light to high-speed information processing. However, the actuation of sound waves in Si PICs usually requires integration of a piezoelectric thin film. Lead Zirconate Titanate (PZT) is a promising  material due to its strong piezoelectric and electromechanical coupling coefficient. Unfortunately, the traditional methods to grow PZT on Silicon are detrimental for photonic applications due to the presence of an optical lossy intermediate layer. In this work, we report integration of a high quality PZT thin film on a Silicon-on-insulator (SOI) photonic chip using an optically transparent buffer layer. 
We demonstrate acousto-optic modulation in Silicon waveguides with the PZT actuated acoustic waves. We fabricate inter digital transducers (IDTs) on the PZT film with a contact photo-lithography and electron-beam lithography to generate the acoustic waves in MHz and GHz range respectively. We obtain a V$_{\pi}$L $\sim$ 3.35 V$\cdot$cm at 576 MHz from a 350 nm thick gold (Au) IDT with 20 finger-pairs. After taking the effect of mass-loading and grating reflection into account, we measured a V$_{\pi}$L $\sim$ 3.60 V$\cdot$cm at 2 GHz from a 100 nm thick Aluminum (Al) IDT consisting of only 4 finger-pairs.  Thus, without patterning the PZT film nor suspending the device, we obtained figures-of-merit comparable to state-of-the-art modulators based on SOI, making it a promising candidate for broadband and efficient acousto-optic modulator for future integration.

\end{abstract}



\section{Introduction}
 {Light-sound interaction in integrated waveguide systems has enabled a wide-range of photonic applications ranging from microwave photonics filters \cite{Marpaung2013,marpaung2015low}, isolators \cite{kang2011reconfigurable, isolator_yu2009, isolator_huang2016}, modulators \cite{nonrecipmodulator_sohn2019, modulator_fan2016integrated, modulator_tadesse2014sub, modulator_tadesse2015acousto, modulator_balram2017acousto, modulator_de2006compact, LiNbO3_Hassanien2021,LiNbO3_Sun2021}, mode-shifters \cite{modeshift_kuhn1971, modeshift_sasaki1974, modeshift_ohmachi1977}, non-reciprocal light transmission \cite{phonon-photon_Safavi19, timreversal_Bahl2018, nonreciprocalBrillouin_Rakich2018,SAW_AlNSOI_Kittlaus2021} and frequency comb generation \cite{freqcomb_Jessen92, freqcomb_Carlos19, freqcomb_Loncar20} to quantum state control and quantum information processing \cite{RevModPhys_kippenberg2014, andrews2014bidirectional,verhagen2012quantum, QAO_Nysten_2017, GaAs_kapfinger2015dynamic}. These devices harness the large overlap between the tightly confined light in a nanophotonic waveguide and acoustic phonons associated with a  mechanical vibration.  
Demonstrations of such acousto-optic interaction in LiNbO$_3$ \cite{LiNbO3_Cai2019, LiNbO3micro2opto_Loncar2019,LiNbO3_Safavi2020, LiNbO3_Hassanien2021, LiNbO3_Sun2021}, GaAs \cite{compactmziGaAs_delima2006,aom_delima2007, GaAs_Kartik2017, GaAs_kapfinger2015dynamic} and InP \cite{InP_Sun95,InP1_Makles2015} based integrated photonics platforms have been reported over the last years. However, on the silicon photonics platform, progress in this field has been lagging as silicon does not exhibit a piezo-electric effect.
Overcoming this deficiency would be a stepping stone towards realizing novel photonic applications,  as the silicon photonics platform is rapidly gaining maturity and already offers a remarkable range of high-performance building blocks, including  modulators, filters, isolators, detectors and lasers. Moreover, its compatibility with CMOS technology offers a route towards mass manufacturing and commercialization \cite{Bahram_Siphotonic2006, SiCMOS_Bogaerts2005}. 
 
 Recently,  thermo-elastic actuation of acoustic waves on SOI was reported \cite{sawthermo1-Avi2019,sawthermo2-Avi2021}. However such a scheme is power-hungry, due to the need of a high power modulated pump source, and costly as they need separate electro-optic modulators. Therefore, a more promising route involves the direct integration of piezoelectric materials on the SOI platform. For instance,  AlN has been integrated with silicon waveguides \cite{AlNforSOI_Tang2012} and used to demonstrate an electrically driven acousto-optic modulator on SOI \cite{SAW_AlNSOI_Kittlaus2021}. This inspired the investigation of other materials exhibiting a strong piezoelectric effect that can be integrated with Si PICs.

Lead Zirconate Titanate (PZT) is one of the most widely used piezoelectric ceramic materials due to its strong piezoelectricity, high electromechanical coupling coefficient, common availability and high temperature compatibility (Curie point $\sim$ 370$^\circ$C) \cite{auld1, piezoreview_2019, piezoceramic_1958}. Hence, integration of a highly textured PZT thin film on SOI could be a promising alternative for realizing efficient optomechanical interactions in Si PICs. However, PZT thin films have traditionally been grown using a Pt seed layer, making them incompatible with photonic technology due to the high optical loss \cite{lossyPZT2017}. Although PZT films sandwiched between two Pt electrodes have been successfully used to obtain stress-optic phase modulation \cite{hosseini2015}, the modulation speed was limited to about 5 kHz. 

Recently, a novel approach for integrating highly textured PZT-films has been developed, using a thin and transparent lanthanide based buffer layer \cite{pzt_George2015}. The high quality of this PZT film has been proven through the demonstration of electro-optic modulators with very low optical loss (<1dB/cm) on both SiN \cite{Alexander2018} and Si photonic platforms \cite{Gilles_EO2020}. 
Additionally, we also demonstrated that this film exhibits a  high second-order nonlinearity ($\chi^2_{zzz} ~128$pm/V) \cite{Gilles_nonlinearPZT2021}. Therefore, a detailed investigation of the piezoelectric effect and acousto-optic modulation with this PZT film is warranted \cite{sawpzt_cleo2020, pztAOsim_Klaus2019, GHzPZT_Irfan2021}.

 In this work, we give a brief theoretical description of the acousto-optic interaction in a waveguide, followed by simulation of the acoustic wave actuation with an inter-digital transducer (IDT). Then we experimentally show the piezoelectric actuation of a surface acoustic wave (SAW) by an IDT fabricated on a PZT-on-glass substrate. Finally, we integrate the PZT thin film on an SOI photonic chip (as shown in figure \ref{fig:idt3D}) and demonstrate acousto-optic modulation in a Si waveguide using an Au IDT . We thereafter present a computational analysis on the effect of the IDT thickness made of a dense metal Au on the transduction efficiency. In the next progression, we use a thinner and lighter metal Al IDT with a smaller period, and demonstrate GHz light modulation with a figure-of-merit comparable to state-of-the-art electro-optic \cite{Alexander2018} and acousto-optic modulators \cite{aom_delima2007,delima_sawmod2013,SAW_AlNSOI_Kittlaus2021}. 
Thus we report the first acousto-optic modulator realised through the direct integration of a photonic compatible PZT thin film on the SOI platform.

\begin{figure}[htpb] 
\centering
{\includegraphics[width=0.5\linewidth]{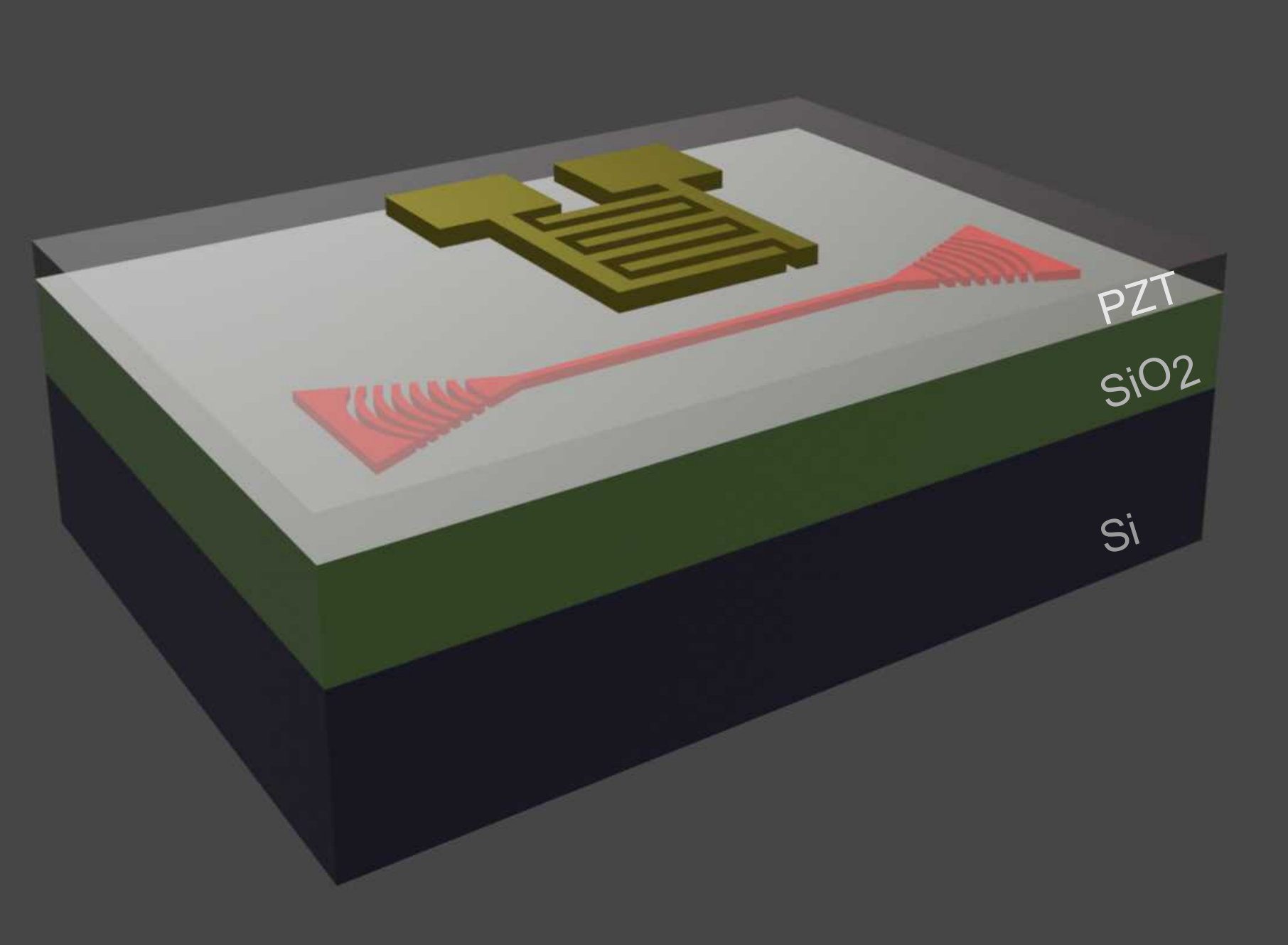}}
\caption{3D drawing of a typical IDT fabricated on an SOI photonic chip. The Si waveguide is embedded between the PZT layer and the SiO$_2$ cladding} \label{fig:idt3D}
\end{figure}

\section{Acoustic wave generation} \label{sec:IDT_actuation}
A surface acoustic wave (SAW) is launched by creating a time-varying periodic deformation pattern along the surface of an elastic material \cite{rayleigh}. 
The propagating SAW is confined near the surface in a region with an extent in the order of its wavelength.
The wavelength of the primary SAW mode is equal to the grating period and its resonance frequency is determined by the acoustic phase velocity of the medium and the boundary condition (free surface) \cite{sawidt1977,sawmat}.  

Thermo-elastic actuation of a SAW has been shown  \cite{sawthermo_0,sawthermo_1,sawthermo_2, sawthermo1-Avi2019, sawthermo2-Avi2021} but in general electro-mechanical actuation \cite{sawdesign} is preferred due to its convenience and efficieny. 
In this method, an IDT consisting of a periodic electrode pattern is deposited on a piezoelectric material.  The electrode-pairs in the IDT allow to generate alternate electric fields in the piezoelectric material, which then results in a periodic strain profile. 
Upon actuation of a piezoelectric material with an IDT, also leaky-SAW modes can be actuated. However this type of waves suffer more from viscous damping in the bulk medium. The leaky-SAW resonance frequency is determined by the IDT-grating period and the longitudinal bulk acoustic velocity of the elastic medium. For a given IDT period, this resonance frequency is usually higher than the SAW resonance frequency (as the longitudinal bulk velocity $>$ SAW velocity) \cite{royerbook2,sawidt1977, 30GHz_leakySAWlinbo3,leakySAWquartz}.   

\section{Simulation of the acoustic wave actuation in PZT} \label{sec:acoustic_simulation} 
The macroscopic behaviour of a polycrystalline PZT film is determined by the net crystallographic orientation of its domains \cite{PZTdomain_Kratzer2018}. The polarization of these domains can be aligned by applying a sufficiently high electric field (the poling process).

\begin{figure}[htpb] 
\centering
{\includegraphics[width=0.6\linewidth]{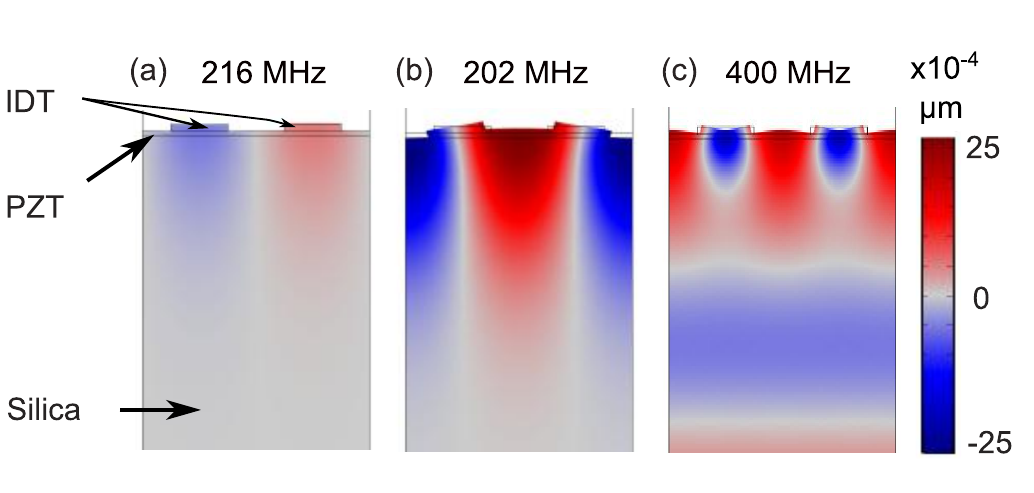}}
\caption{A 2D finite element method (FEM) simulation shows the vertical displacement from the SAW actuation for three PZT domain polarities: (a) out of the substrate plane, (b) uniformly in-plane, perpendicular to the IDT fingers, and (c) periodically oriented along the IDT electric field lines.} \label{fig:ComsolfigPolingscemes_panel}
\end{figure}

In figure \ref{fig:ComsolfigPolingscemes_panel}, we show how the electromechanical actuation depends on the PZT domain polarity. We consider three different configuraitons: (a) out-of-plane polariztion, (b) uniform in-plane polarization, perpendicular to the IDT-electrodes, and (c) polarization periodically oriented along the electric field lines applied by the IDT. We notice that the resulting displacement is much weaker with the out-of-plane oriented PZT domains (a), compared to the in-plane oriented PZT domains (b),(c). This can be explained from the inverse piezoelectric tensor for PZT. From the piezoelectric constitutive equation, the induced strain is determined as \cite{piezoIEEEstd},  
\[
    \begin{bmatrix}
        S_1 \\
        S_2 \\ 
        S_3 \\ 
        S_4 \\ 
        S_5 \\ 
        S_6 
    \end{bmatrix} \quad =
    \begin{bmatrix}
        0 & 0 & d_{31}  \\
        0 & 0 & d_{31}  \\
        0 & 0 & d_{33}  \\
        0 & d_{24} & 0  \\
        d_{15} & 0 & 0  \\
        0 & 0 & 0  
    \end{bmatrix}
    \begin{bmatrix}
        E_x \\
        E_y \\
        E_z 
    \end{bmatrix},
\]
where $d_{ij}$ is the tensor describing the inverse piezoelectric effect for PZT and $E_i$ is the applied electric field. In the simulation, we used the following inverse-piezoelectric coefficients for PZT: $d_{31}$ =-171 pm/V, $d_{33}$ =374 pm/V, $d_{15} = d_{24} =$ 584 pm/V \cite{pzt5a}.

From this relation we see that for a longitudinal actuation of PZT (necessary to launch a SAW), the applied electric field should be along the $z$-axis ($E_z$) in the crystal coordinate system, i.e. along the PZT domain polarization.
Therefore, in order to have an effective SAW actuation with an IDT, the domains should be aligned (uniformly or periodically) in the substrate plane, perpendicular to the IDT-electrodes. 

Furthermore, we notice that a periodic orientation of the PZT domains (figure \ref{fig:ComsolfigPolingscemes_panel}(c)) results in a SAW resonance frequency (400 MHz) that is almost twice the SAW resonance frequency (202 MHz) from the PZT with all domains uniformly in-plane oriented (figure \ref{fig:ComsolfigPolingscemes_panel}(b)). This is because the actuation with an periodic domain polarity results in a SAW wavelength that is half of the IDT period, hence the resonance frequency doubles.

\section{Acousto-optic interaction}\label{sec:ao_formulation} 
Next, we consider a PZT film integrated on an SOI photonic chip as schematically illustrated in figure \ref{fig:idt3D}.
The acoustic wave excited by the IDT results in a dynamic strain profile, which perturbs the refractive index of the medium (photo-elastic effect). In the waveguide, this index modulation diffracts the incident carrier mode into two sidebands.     
After travelling through the modulator of length $L$, the modulated field can be described as (for detail, see the appendix),

\begin{equation}  \label{eqn:DUTout}
   \psi_1(t)  = \Re A_1\big\{e^{i\omega_0 t} +  \frac{\alpha(L)}{2} [ e^{i(\omega_0 +\Omega)t} - e^{i(\omega_0 -\Omega)t} ] \big\}       
\end{equation}

We thereby assumed that the modulator is short, thus the phase mismatch between the modulated signals (sidebands) and the carrier mode is negligible. 

In this equation (\ref{eqn:DUTout}), $A_1$ is the electric field amplitude of the output light from the DUT (waveguide), $\omega_0$ is the angular frequency of the incident light, $\Omega$ is the angular frequency of the acoustic wave and $\alpha(L)$ is the phase modulation amplitude. $ \alpha(L) = -(2\pi \Delta n_{eff} L/\lambda_0) $ where $\Delta n_{eff}$ is the effective index change induced by the acoustic wave and $\lambda_0$ is the free space wavelength of the light. $\alpha(L)$ is a measure of the modulation efficiency and depends on the coupling between the induced acoustic field and the optical field in the waveguide. 

\begin{figure*}[h!]
\centering
{\includegraphics[width=0.9\linewidth]{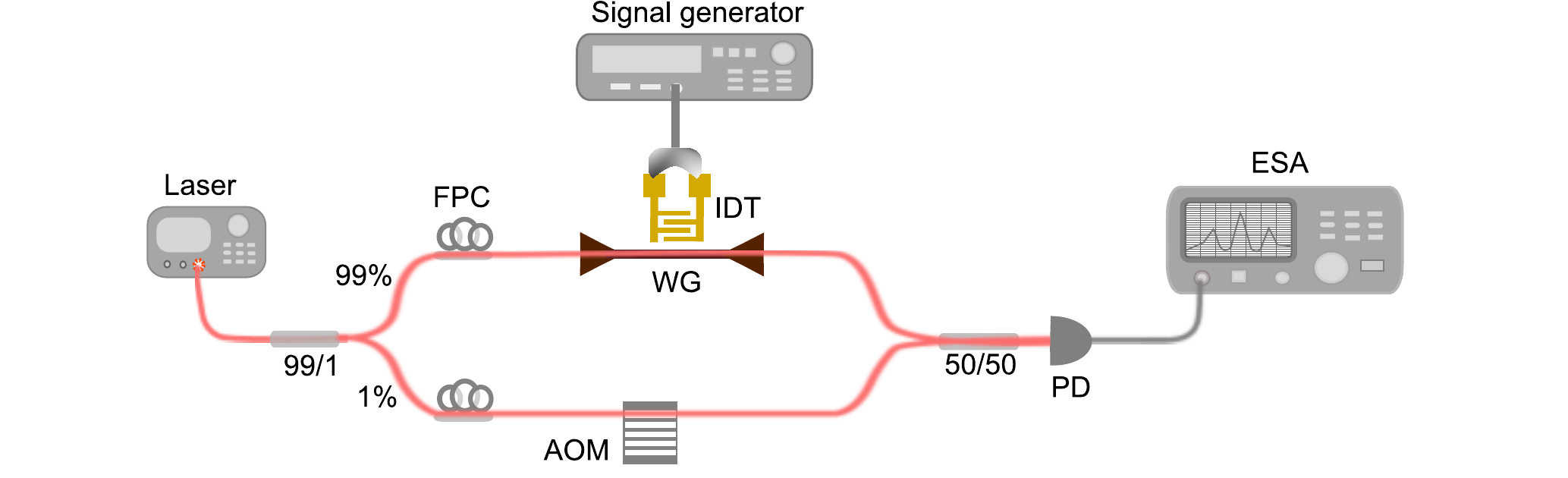}}
\caption{Schematic of the heterodyne setup used to measure the acousto-optic phase modulation. A commercial acousto-optic modulator (AOM) was used as a local oscillator to frequency shift the reference signal by 200 MHz. This frequency shifted signal was then mixed with the phase modulated signal emerging of the DUT via a 3dB fiber combiner to down-convert the carrier signal on a detector. } 

\label{fig:deviceSetup_pane1}
\end{figure*}
To experimentally measure the optical phase modulation, we used a heterodyne setup as shown in figure \ref{fig:deviceSetup_pane1}. The output signal of the modulator is combined with a frequency-shifted signal from an acousto-optic modulator(AOM) to down-convert the carrier signal in the photodetector. 
The frequency shifted signal exiting the AOM is given by:
\begin{equation}
   \psi_2(t)  =  \Re A_2\big\{e^{i(\omega_0 +\Delta\omega)t} \big\}  \label{eqn:AOMout}
\end{equation}
Where $A_2$ is the amplitude of the output light from AOM and $\Delta\omega =2\pi \times$ 200 MHz is the frequency shift induced by the AOM. Thus, the field coupled into the photodiode (PD) is:
\[
 \psi_{PD}(t) = \psi_1(t) + \psi_2(t)   \]
The output current from the PD, $I_{PD}$ is equal to the responsivity of the PD times the optical input power  $\implies I_{PD} \propto  |\psi_{PD}|^2 $. The electrical power measured by the ESA equals $P_{ESA} =  I^2_{PD} \times Z_0 $, where $Z_0$ is the load impedance (50 $\Omega$).  Hence, the power spectrum measured in the ESA is given by: 
\begin{equation}
 \begin{split}   
    P_{ESA}(t) &\propto  \big[ |A_1|^2 + |A_2|^2 + (A^*_1A_2+A^*_2A_1)\big[ cos(\Delta\omega t) \\
    &+\frac{\alpha(L)}{2} \big\{ cos((\Delta\omega +\Omega)t)-cos((\Delta\omega -\Omega)t) \big\}  \big]  +|A_1|^2 \alpha(L)^2sin^2(\Omega t)  \big]^2
\end{split}
\end{equation}
We can ignore the last term, as $\alpha(L)<<$ 1, and unmodulated terms contributing only to the DC current. 
We then obtain,
\begin{equation}    \label{eqn:ESAout}
\begin{split}
 P_{ESA}(t) \propto 4|A_1|^2|A_2|^2 \bigg[cos(\Delta\omega t) +  \frac{\alpha(L)}{2} \Big\{ cos((\Delta\omega+\Omega)t)-cos((\Delta\omega-\Omega)t) \Big\}  \bigg]^2
\end{split}
\end{equation}
The Fourier transform of  equation (\ref{eqn:ESAout}) gives a peak signal at the AOM frequency shift $\Delta\omega$, and two sideband peaks at frequencies $|\Delta\omega \pm \Omega|$ introduced by the acoustic wave modulation.
Further algebra on equation (\ref{eqn:ESAout}) gives us the following expression for the modulation amplitude $\alpha(L)$,
\begin{equation} \label{eqn:betafrompeaks}
\begin{split}
P^{\textrm{AOM}}_{ESA} [dBm]-P^{\textrm{Sideband}}_{ESA} [dBm] 
=20 \, \textrm{log} \Big[\frac{\alpha(L) [\textrm{rad}]}{2}\Big]
\end{split}
\end{equation}
Hence, the modulation amplitude $\alpha(L)$ can be extracted independent of the photodetector responsivity and gain. Now, with this $\alpha(L)$, we can calculate the voltage required for a $\pi$- phase shift $V_{\pi} = \pi \, V_{RF}/\alpha(L)$, where $V_{RF}$ is the voltage amplitude of the input RF signal.

For the power of the upper sideband peak in equation (\ref{eqn:ESAout}) we find: \[
    P^{\textrm{Sideband}}_{ESA} \propto |A_1|^2|A_2|^2 \, \alpha(L)^2 \Big( cos((\Delta\omega+\Omega)t) \Big)  \]
\[ 
    \implies  P^{\textrm{Sideband}}_{ESA} \propto P^2_{laser} P_{RF}   \]
As $\alpha(L)^2 \propto P_{RF}$, where P$_{RF}$ is the applied RF power to the IDT, and $|A_1|^2|A_2|^2  \propto P^2_{laser}$,  where $P_{laser}$ is the input laser power. Thus, the above equation can be written as, 
\begin{equation} \label{eqn:sidebandpower}
\begin{split}
    P^{\textrm{Sideband}}_{ESA} [dBm] =  2P_{laser}[dBm] + P_{RF} [dBm] 
    + constant
\end{split}
\end{equation}
This shows how the modulated power depends on the input laser power and the driving RF power.  

\section{Materials and methods}
We fabricated the PZT thin films by chemical solution deposition (CSD) as described in \cite{johnPZT}.
The ultrathin (5-15 nm) lanthanide-based buffer layer (La$_2$O$_2$CO$_3$) used in this method works as an excellent lattice match, resulting in a uniform, crack-free and preferentially c-oriented growth of the PZT thin film \cite{johnPZT}. Moreover, this intermediate buffer layer provides an efficient diffusion barrier between the substrate and the PZT layer. After spin-coating the buffer solution on a substrate, we annealed the sample at 400-500$^{\circ}$C. Subsequently, we spin-coated the PZT precursor solution and subjected it to pyrolysis at 300$^{\circ}$C (we repeated this cycle to obtain a thicker PZT film). Then we annealed the amorphous PZT layer in a tube furnace at 500-600$^{\circ}$C for 20-30 min in oxygen ambient to let it crystallize. We finally obtained a PZT film with chemical composition PbZr$_{0.52}$Ti$_{0.48}$O$_3$.

\begin{figure*}[h!]
\centering
{\includegraphics[width=0.95\linewidth]{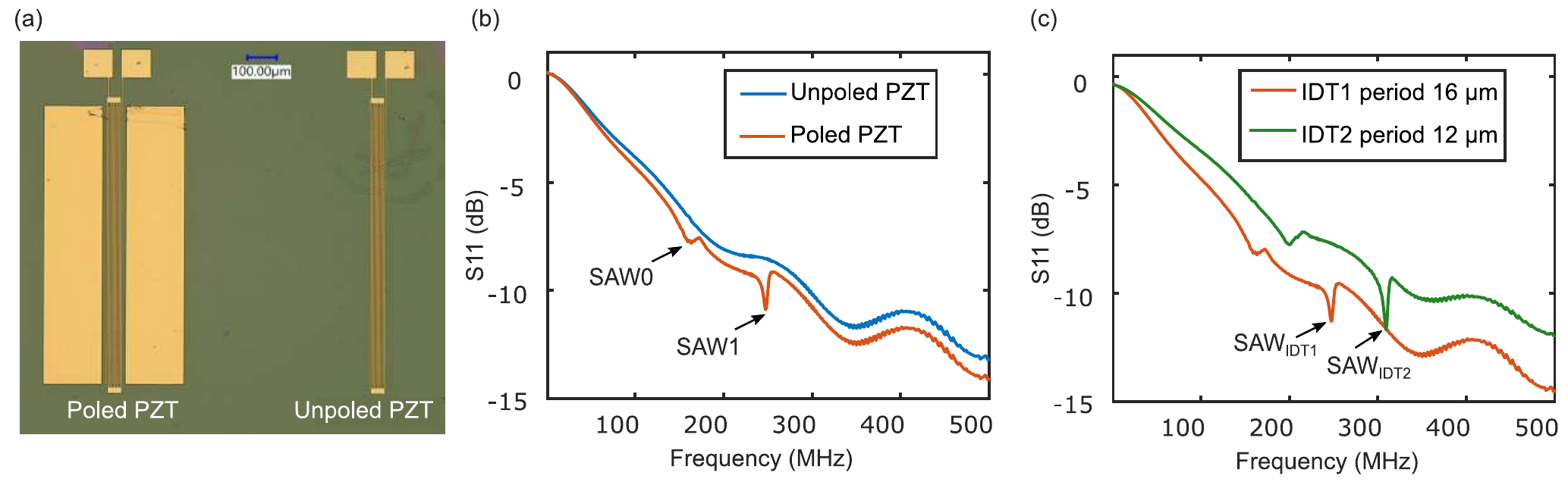}}
\caption{(a) Microscope image showing a test IDT fabricated in between parallel electrode bars (80 $\mu$m spacing) for poling process, and an identical control IDT fabricated without the electrode bars. (b) $S_{11}$ signal measured for test (poled) and control (unpoled) IDT with period 16 $\mu$m. (c) $S_{11}$ signal measured for poled IDTs with period 12 $\mu$m and 16 $\mu$m.
}
\label{fig:singleIDTsaw_panel}
\end{figure*}

To demonstrate the piezoelectric properties of our PZT film, we designed standalone IDT structures on PZT films deposited on a glass substrate.  
In a first fabrication step, we patterned parallel electrodes (separation 80-150 um) with laser direct write lithography (LDW), followed by deposition of a 20 nm Ti/ 350 nm Au layer through thermal evaporation and a lift-off process. 
These metal electrodes were used to uniformly pole the PZT film in-plane, for about 1 hour at 40$^{\circ}$C. The applied DC voltage varied between 820-1100 V, depending on the electrodes separation. 
In the second step, we used the same metallization process for fabricating test IDTs (in the poled PZT region) and control IDTs (in the unpoled region) as shown in figure \ref{fig:singleIDTsaw_panel}(a). The number of finger-pairs was limited to about 5-7 given the resolution of the optical lithography process and the limited spacing between the poling electrodes. 
On these IDTs, we measured the frequency dependent electrical reflection parameter $S_{11}$ using a Fieldfox- vector network analyzer (VNA).

For characterising the excitation of acoustic waves and their interaction with integrated waveguides, we used an SOI photonic chip, fabricated through a multi-project wafer (MPW) run, as the starting point. In the MPW fabrication process, the waveguides were defined in a 220 nm thick silicon layer on top of a 2 $\mu$m buried oxide layer. The chips were planarized using oxide deposition and chemical mechanical polishing (CMP). We deposited a 10-15  nm thick lanthanide-based buffer layer, followed by a 200 nm thick thick PZT-layer on these SOI chips. We then defined IDTs on top of the PZT layer using optical lithography, followed by deposition of 20 nm Ti/ 350 nm Au layer through thermal evaporation and a lift-off process as shown in figure \ref{fig:fabimage}. Contrary to the approach taken for the glass substrate described in the previous paragraph, for these chips the PZT in the IDT region was poled using the IDT itself, by applying a voltage of 30-60 V (depending on the IDT-finger spacing) at 40$^{\circ}$C for about 40 min.

\begin{figure*}[h!]
\centering
    {\includegraphics[width=0.7\linewidth]{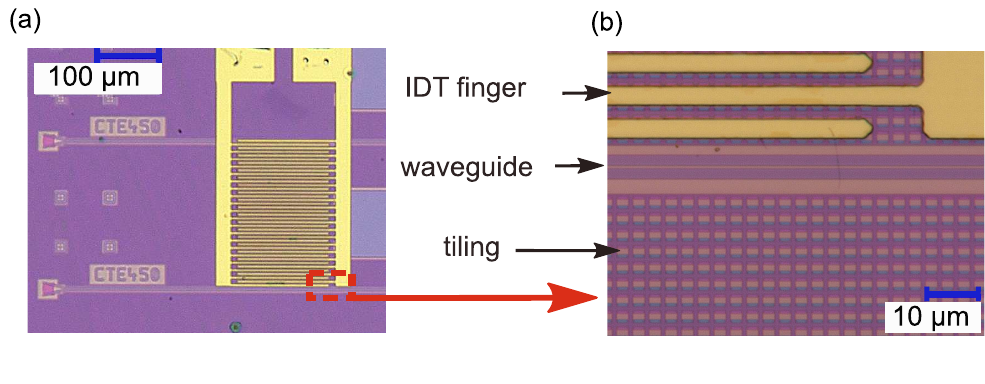}}
    \caption{(a) Microscope image of an IDT with finger-width 3$\mu$m, aperture 180$\mu$m and 20 finger-pairs, (b) zoomed-in image showing the tiling pattern (containing shallow and deep etched Si) distributed over the  regions without waveguide designs. These 220 nm thick patterns have a periodicity of 3 $\mu$m and are defined in the Si-device layer to maintain uniformity during the wafer fabrication.}
    \label{fig:fabimage}
\end{figure*}

The experimental setup used for the measurement of the optical phase modulation is shown in figure \ref{fig:deviceSetup_pane1}. The signal  (1550 nm) from the laser (Santec TSL510) was divided into two parts by a 99/1 fiber splitter, with 1\% of the signal routed through an acousto-optic modulator (Gooch \& Housego Fiber-Q, driver model 1200 AF-AINA-2.5 HCR) to frequency-shift the signal by 200 MHz. The other 99\% of the signal was fed into the DUT. A Rhodes \& Schwarz-SMR40 signal generator was used as an RF source to actuate the IDT. The modulated signal from the waveguide was mixed with the output signal from the AOM via a 50/50 fiber combiner and fed into a fast TIA photodetector (Thorlab PDB480C or PD-40GHz Discovery LabBuddy). The output of the PD was analysed with an electrical spectrum analyzer (Agilent N9010A). 
 

\section{Results and discussions}

In figure  \ref{fig:singleIDTsaw_panel}(a) we show a microscope image of a test IDT (between electrodes) and an identical, control IDT (without electrodes) fabricated on a PZT-on-glass substrate. The PZT film under the test IDT is uniformly in-plane poled (using the parallel electrodes), whereas for the control IDT, it is unpoled. Figure  \ref{fig:singleIDTsaw_panel}(b) shows the electrical reflection parameter $S_{11}$ measured on the two types of IDTs (period 16 $\mu$m). We observe two dips at 161 MHz and 247 MHz in the $S_{11}$ measured on the test IDT and none from the control IDT.    
These dips indicate the actuation of acoustic waves because the input RF power is now converted into the acoustic power, lowering the reflected RF power. 
Therefore, this result already indicates that the IDT deposited on the poled PZT is more efficient in actuating acoustic waves  compared to the IDT on as deposited (unpoled) PZT. 
Figure \ref{fig:singleIDTsaw_panel}(c) shows the $S_{11}$ response for two IDTs with different period.  We observed that the resonance frequencies shift proportionally with the IDT period.  This corroborates that indeed the $S_{11}$ dips are from the acoustic wave actuation as their excitation frequencies are determined by the IDT period. Thus we conclude that our PZT film exhibits a piezo-electric effect and allows for the actuation of acoustic waves.

With the uniform in-plane poling scheme using an electrode-pair across the IDT pattern, the size of the poled PZT region is limited due to constraints on the the maximum applicable voltage. Consequently, the size of the actuators is also limited. Therefore, in subsequent devices, we poled the PZT film using the IDT electrodes themselves.

\begin{figure*}[htpb]
\centering
\includegraphics[width=\linewidth]{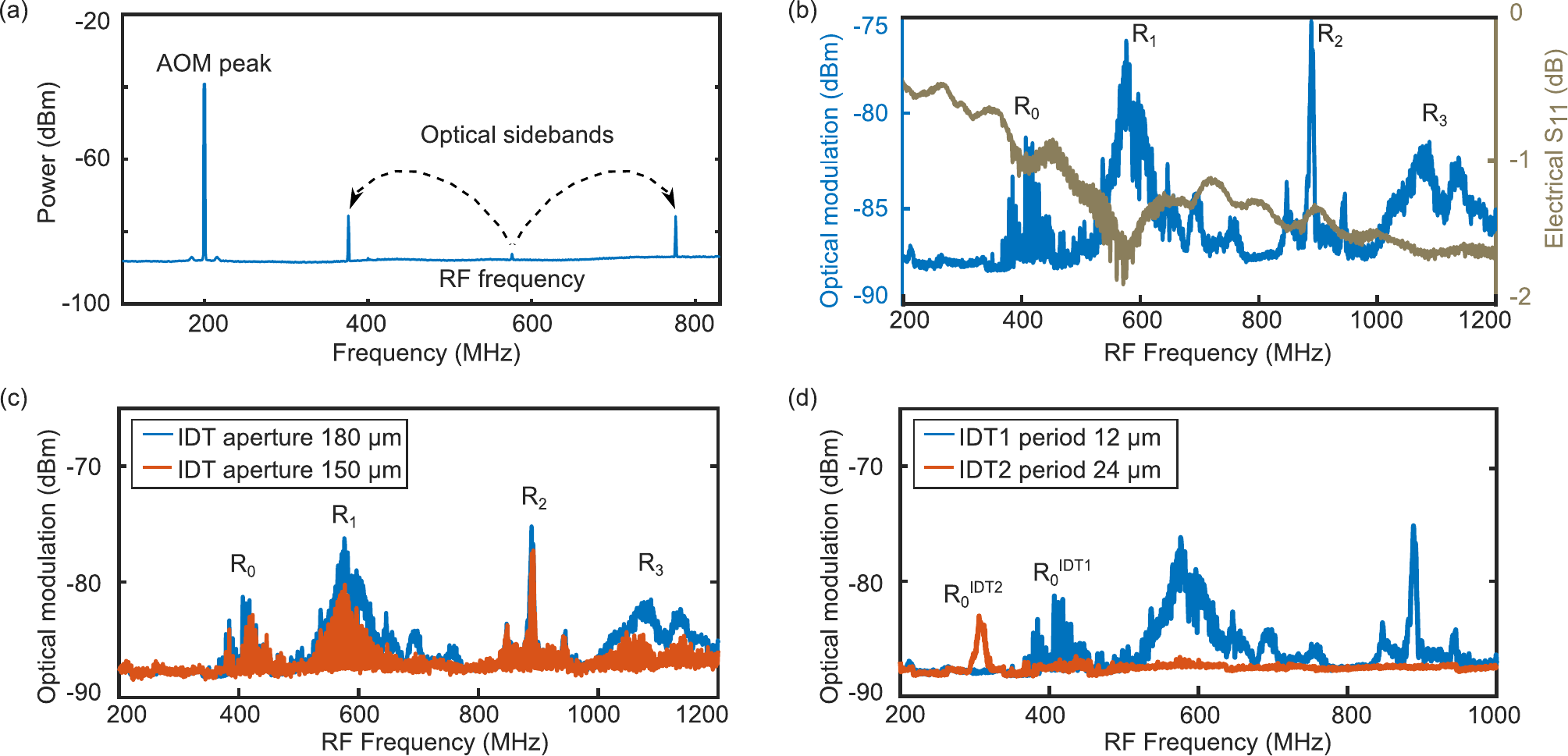}
\caption{(a) Measured output spectrum for IDT with period 12 $\mu$m actuated with a 15 dBm RF signal at 576 MHz, showing the carrier peak at the AOM driving frequency (200 MHz) and optical sidebands at 576 $\pm$ 200 MHz. (b) Frequency dependent modulation spectrum (blue) and $S_{11}$ (brown) for IDT with period 12 $\mu$m.  The $S_{11}$ spectrum shows dips at 410 MHz and 576 MHz, while the measured optical modulation spectrum shows peaks at 410 MHz, 576 MHz, 888 MHz and 1088 MHz. (c) Modulation spectra for IDT with aperture 180 $\mu$m and 150 $\mu$m.  (d) Modulation spectra for IDT1 (with period of 12 $\mu$m, aperture 180 $\mu$m, 20 finger-pairs), placed 6 $\mu$m away from the waveguide, and IDT2 (with period 24 $\mu$m, aperture 280 $\mu$m and 30 finger-pairs), placed 1024 $\mu$m away from the waveguide.}
\label{fig:OptMod_panel}
\end{figure*}

Next, we characterised the strength of the acousto-optic interaction, using an SOI photonic chip with integrated PZT film.
Figure \ref{fig:OptMod_panel}(a) shows the output power spectrum when an IDT with period 12 $\mu$m is actuated with a 15 dBm RF signal at 576 MHz. As described in equation \ref{eqn:ESAout}, the peak at 200 MHz corresponds to the AOM driving frequency and the two sidebands of the carrier, at 576 $\pm$ 200 MHz, result from the acousto-optic modulation. To characterize the modulation strength, we measured the amplitude of the upper sideband peak with respect to the applied RF frequency, for different IDTs, as shown in figure \ref{fig:OptMod_panel}(b), (c) and (d). 

In figure \ref{fig:OptMod_panel}(b), we compare the results from the optical modulation measured in the waveguide and the electrical $S_{11}$ measured on the corresponding IDT (with period 12 $\mu$m and aperture 180 $\mu$m). 
We notice that the optical modulation peaks at 410 MHz and 576 MHz, are consistent with the transduction dips in the electrical $S_{11}$ response. For the modulation peaks at 888 MHz and 1088 MHz, however, we do not see clear dips in the $S_{11}$ signal. We suspect this might be due to the additional noise or cross-talk at higher RF frequencies in the $S_{11}$ measurement. Thus, both the electrical and optical measurements confirm the excitation of acoustic waves and corresponding optical phase modulation in the waveguide. 

\begin{figure*}[ht]
\centering
\includegraphics[width=0.95\linewidth]{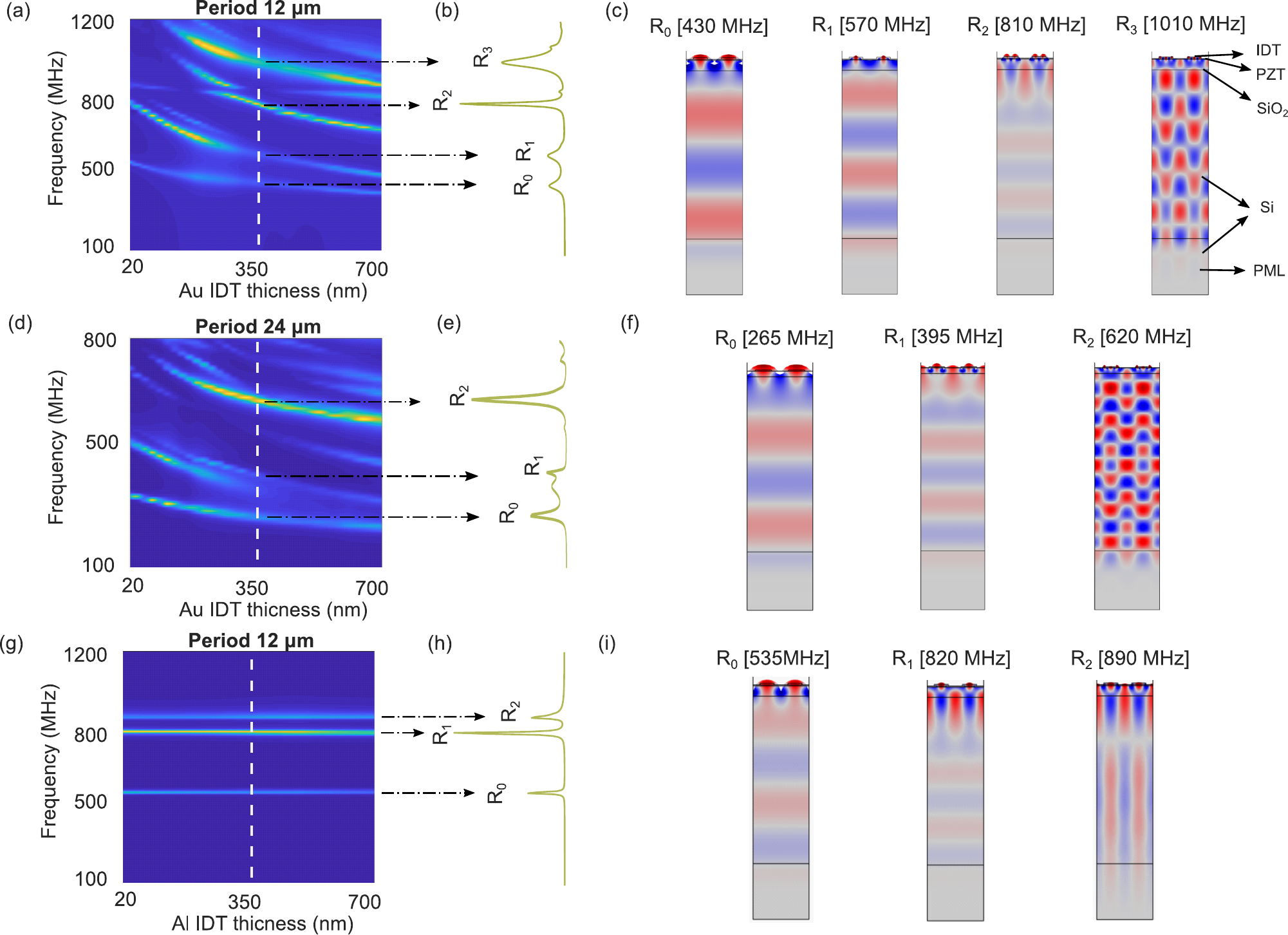}
\caption{FEM calculated acoustic dispersion diagram with respect to the IDT thickness from a unit cell of (a) 12 $\mu$m period Au IDT, (d) 24 $\mu$m period Au IDT and (g) 12 $\mu$m period Al IDT. The color map represents normalized elastic strain energy. (b), (e) and  (h) show the elastic strain energy spectrum plotted for 350 nm IDT thickness. (c), (f) and (i) show the vertical-displacement profile of the corresponding modes for a 350 nm thick IDT. For the simulation details see the appendix.}
\label{fig:Dispersion_panel}
\end{figure*}

In figure \ref{fig:OptMod_panel}(c), we show the modulation spectra from devices with IDT aperture 180 $\mu$m and 150 $\mu$m. As expected, we see a proportional increment in the modulation strength when the modulation length (IDT aperture) is increased. In figure \ref{fig:OptMod_panel}(d), we show the modulation spectra from devices with two different IDT periods. The spacing between the IDT aperture and the WG core edge is 6 $\mu$m for IDT1 (period 12 $\mu$m, No. of finger-pairs 20) and 1204 $\mu$m for IDT2 (period 24 $\mu$m, No. of finger-pairs 30). In the latter case (IDT2), the modulation spectrum shows only one peak while for IDT1 multiple peaks are visible. To understand this and get insight into the characteristics of the acoustic modes corresponding to these peaks, we carried out a detailed FEM simulation on a unit cell of the IDT, assuming periodic boundary conditions.  The details of the simulation setup are discussed in the appendix).
Figure \ref{fig:Dispersion_panel} shows the main results. In figure \ref{fig:Dispersion_panel}(a) we show the acoustic mode dispersion with respect to the thickness of a Au IDT (period 12 $\mu$m). Figure \ref{fig:Dispersion_panel}(b) shows the elastic energy spectrum for a thickness of 350 nm indicating 4 peaks, and the associated mode profiles shown in figure \ref{fig:Dispersion_panel}(c). We note that R$_0$ at 430 MHz, R$_1$ at 570 MHz, R$_2$ at 810 MHz and R$_2$ at 1010 MHz correspond to the fundamental SAW mode, a pseudo-SAW mode, a leaky-SAW mode and a leaky mode. These can be linked to the peaks observed in the experimental spectrum shown in figure \ref{fig:OptMod_panel}(c) at 410 MHz, 576 MHz, 888 MHz and 1088 MHz for IDT1. 
The small deviation between experimental and simulated results can be attributed to a slight mismatch in the geometric and material parameters used in the simulation and the real experiment. Additionally, the Si tiling pattern on the chip (see figure \ref{fig:fabimage}) was not included into the simulation.

Similarly, figure \ref{fig:Dispersion_panel}(f), where the period was increased to 24 $\mu$m, shows the excitation of 3 main modes, whereby R$_0$ at 265 MHz, R$_1$ at $\sim$ 395 MHz and R$_2$ at 620 MHz correspond to the fundamental SAW mode, a pseudo-leaky SAW mode, and a leaky mode respectively. Given the leaky character of R$_1$ and R$_2$, only R$_0$ can travel a longer distance, explaining why we observed only a single modulation peak in figure \ref{fig:OptMod_panel}(d) for IDT2.

From the acoustic dispersion diagrams shown in figure \ref{fig:Dispersion_panel} (a) and (d), we notice that when the IDT thickness is negligible, there are three main modes excited: the fundamental SAW mode, a leaky-SAW mode and a higher order (weak) SAW mode. However, when we increase the IDT thickness, the added mass-loading and grating reflection decreases the mode frequencies and introduces more leaky modes. 
While the mass-loading effect from the IDT can be desirable for some sensing applications \cite{massloading_chen2020}, in our case, it is undesirable as it weakens the SAW transduction strength and diffracts more energy into the substrate. As a result, the acousto-optic interaction in a WG in the surface layer gets weaker. To solve the problem of grating reflection, in some cases a split-finger IDT is used\cite{splitfinger-IDT_delima2013, royerbook2}, because the reflected waves from the grating then add up destructively. However in case of periodic poling, this condition isn't satisfied.  Therefore, in our case, thinner and lower density metal electrodes could be a possible solution. 

In figure \ref{fig:Dispersion_panel}(g), we show the acoustic dispersion diagram for an Aluminum IDT, which has a low density (density $\sim$ 1/7 of Au) and lower acoustic impedance mismatch. Now we do not see any shift in the mode frequencies.

Therefore, in the next progression of fabrication, we used Al for designing the IDT. We used electron-beam lithography to pattern a small period IDT, which allows the excitation of higher (GHz) frequency acoustic waves. We deposited 100 nm thick Al using e-gun evaporation, followed by a metal-lift off process. Then we patterned additional contact pads with photo-lithography, followed by a $\sim$ 330 nm thick Al deposition and lift-off process. Finally, we covered the top of the device with a 50 nm Al$_2$O$_3$ layer using atomic-layer deposition (ALD). An example of the resulting devices is shown in the inset of figure \ref{fig:GHzmodulation}.  

\begin{figure}
\centering
{\includegraphics[width=0.5\linewidth]{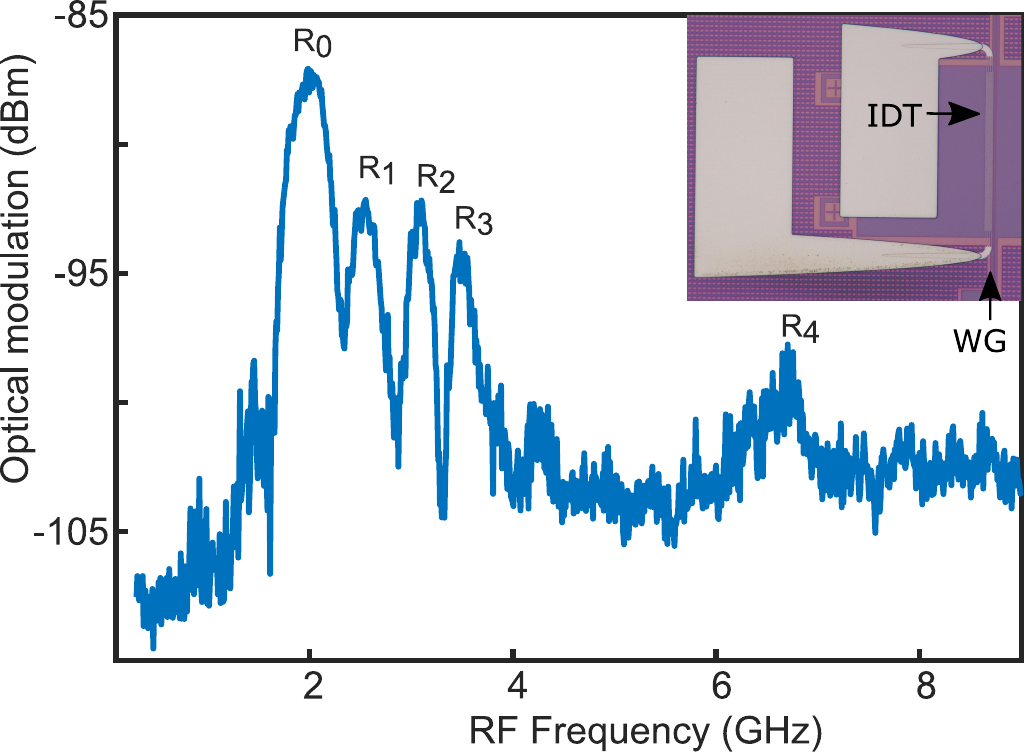}}
\caption{Measured optical modulation spectrum for an Al IDT with 4 finger-pairs, period 2 $\mu$m (finger-width 500 nm) and aperture length 70 $\mu$m, actuated with 12 dBm RF power.  The spectrum shows modulation peaks at 2 GHz (R$_0$), 2.55 GHz (R$_1$), 3.1 GHz (R$_2$), 3.47 GHz (R$_3$) and 6.7 GHz (R$_4$).
The inset shows a microscope image of the fabricated device, which was spaced 2 $\mu$m from the WG in a region without Si tiling.} \label{fig:GHzmodulation}
\end{figure}
   
Figure \ref{fig:GHzmodulation} shows the modulation spectrum measured from a 100 nm thick Al IDT with period 2 $\mu$m driven with 12 dBm of RF power. We find modulation peaks at 2 GHz (R$_0$), 2.5 GHz  (R$_1$), 3.1 GHz (R$_2$), 3.5 GHz (R$_3$) and 6.7 GHz (R$_4$). From an FEM simulation, we conclude the first and the strongest modulation peak (R$_0$) to be the fundamental SAW mode \cite{GHzPZT_Irfan2021} while the other peaks belong to higher order modes.

In figure \ref{fig:power} we show how the modulated power depends on the input laser power and the RF driving power. These measurements were done on the IDT of period 12 $\mu$m at an RF frequency of 576 MHz. The slopes obtained from a linear fit of the data are very close to the expected values as discussed in equation \ref{eqn:sidebandpower}. The small deviation is attributed mainly to the higher order terms in the phase modulation, which are not negligible anymore at higher RF power, and poor signal to noise ratio at lower optical power. 

\begin{figure}[ht!]
\centering
    {\includegraphics[width=0.45\linewidth]{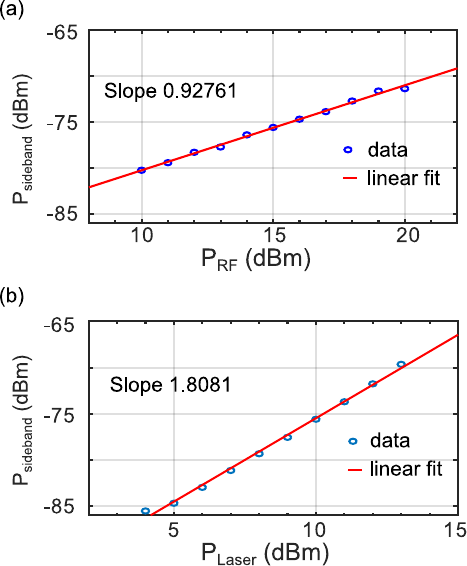}}
    \caption{ Modulated power (sideband peak) dependence on, (a) the driving RF power at a fixed laser input of 10 dBm, and (b) the laser power at a fixed RF power of 15 dBm. The measurement was done on the IDT of period 12 $\mu$m at an RF frequency of 576 MHz}
    \label{fig:power}
\end{figure}

From comparing the AOM peak (-39.15 dBm) and the modulation peak (-75.60 dBm) in figure \ref{fig:OptMod_panel}(a), we extract the phase modulation amplitude $\alpha(L)$ using equation \ref{eqn:betafrompeaks}. We obtained $\alpha(L)$ to be $\approx$ 0.03 $rad$, when the Au IDT with period 12 $\mu$m, 20 finger-pairs and aperture of 180 $\mu$m is actuated with 15 dBm RF power at 576 MHz. The corresponding $V_{\pi}$L is $\approx$ 3.35 V$\cdot$cm. Similarly, for the Al IDT of period 2 $\mu$m and aperture 70 $\mu$m, the modulation peak at 2 GHz (R$_0$) was measured to be at -87.3 dBm (figure \ref{fig:GHzmodulation}), while the AOM peak was noted at -43.5 dBm . This gives $\alpha(L)$ $\approx$ 0.0077 $rad$ and the corresponding $V_{\pi}$L to be $\approx$ 3.60 V$\cdot$cm. This is a significant improvement considering that the Al IDT now consists of only 4 finger-pairs. While we believe the main reason for this improvement is a low mass-loading and grating reflection from the Al IDT as discussed above, the other contributing factors could be the absence of the Si tiling pattern and a stronger acousto-optic overlap in the waveguide (width 0.45 $\mu$m) from the acoustic wave (wavelength $\sim$ $1\mu$m).

With a similar PZT layer, a modulator exploiting the electro-optic (EO) effect has been reported to exhibit $V_\pi$L $\approx$ 3.2 V$\cdot$cm \cite{Alexander2018}.
In \cite{SAW_AlNSOI_Kittlaus2021}, using a Si waveguide integrated with AlN and an IDT consisting of 107 finger-pairs actuated at 3.11 GHz, the $V_\pi$L was reported to be 1.8 V$\cdot$cm.
Thus, our figures of merit are competitive to state-of-the-art electro-optic and acousto-optic modulators integrated on Si PICs

\section*{Conclusion}
We investigated the piezo-electrical actuation of surface acoustic waves using a photonic compatible PZT film.
Then we integrated a PZT thin film on a planarized SOI photonic chip to induce acousto-optic modulation in a waveguide with MHz acoustic waves from the Au IDT. 
Thereafter, through FEM analysis, we pointed out the issue of mass-loading and grating reflection from the Au IDT. We then fabricated a new device with an Al IDT and smaller period to actuate GHz acoustic waves. We obtained a competitive $V_\pi$L $\approx$ 3.6 V$\cdot$cm with an Al IDT consisting of only 4 finger-pairs, without patterning the PZT layer or under-etching the device. 
Further improvements in the device performance are expected when the scattering and damping loss factors are eliminated. For instance, 
the IDT design can be optimized to match the electrical impedance, thus minimise the RF power reflection. Additionally, the current bi-directional IDT actuates the acoustic waves in both directions, thus only half of the acoustic energy is utilised for the modulation. We can define an acoustic reflector to collect the other half or design a unidirectional SAW actuator \cite{uniSAW1, uniSAW2}. Furthermore, to avoid leakage of any acoustic energy into the substrate and any interference from the bulk acoustic waves, the device could be under-etched. 
 
Thus, we demonstrated that our PZT-film, deposited on planarised silicon photonics chips, exhibits a strong piezoelectric effect and can be exploited to achieve strong phonon-photon coupling in microscale waveguides.
Through this hybrid integration process, we hope to realize power efficient, miniaturized and scalable piezoelectric micro-actuators based photonic components such as tunable filters, isolators, modulators, switches and beam-steering.

\section{Appendix}

\subsection{Elastic wave actuation}   \label{appen:elastic}
The linear piezoelectric constitutive equations (in strain-form) are given by \cite{piezoIEEEstd},
\begin{equation}
\begin{split}
    D = d T + \epsilon_T E  \\
    S = s_ET + d^TE
\end{split}   
\end{equation}
where $D$ is the displacement field, $d$ is the inverse piezoelectric tensor, $T$ is the applied stress, $\epsilon_T$ is the permittivity tensor, $E$ is the applied electric field, $S$ is the strain tensor and $s_E$ is the elastic compliance (inverse of stiffness/elastic) tensor of the material.

These piezoelectric constitutive equations, along with a Newtonian equation of motion are solved to obtain the electromechanical dynamics of the acoustic modes. For a multilayer substrate with an anisotropic piezoelectric layer, these solutions are usually obtained numerically. We used a commercial FEM solver, COMSOL Multiphysics 5.5 extensively to simulate the piezoelectric actuation and the resulting mechanical displacement profiles. 

To accurately simulate the case of periodic poling of the PZT layer by the IDT electrodes, first we calculated the electric field using an electrostatic simulation.
Then in the second step, the PZT domain orientation was aligned along these electric field lines. Now with this setup, we carried out an FEM simulation in the frequency domain to obtain the response to an RF actuation (amplitude 1V). 

For the FEM simulations shown in figures 2 and 7, we applied a Floquet periodic boundary condition on the left and right boundaries of the unit cell. In the simulation shown in figure 2, we defined a 350 nm thick Au IDT of period 12 $\mu$m  on a 200 nm thick PZT film on a silica substrate of thickness 4 $\times$ IDT period. And for the simulation shown in figure 7, we defined the IDTs on a multi-layered structure consisting of a 200 nm PZT layer,  a 2.220 $\mu$m SiO$_2$ layer and a Si substrate of thickness 4 $\times$ IDT period. We set the bottom 1 period of the unit cell as perfectly matched layer (PML) domain and the bottom-most boundary as low-reflecting boundary to absorb the incoming acoustic waves. 

To visualise the propagation of the IDT actuated acoustic waves and calculate the acousto-optic interaction in the waveguide, we created a 2D simulation setup as shown in figure \ref{fig:SAWthicknessEffectPML}(a). We extended the substrate region beyond the IDT and terminated it with a PML layer (thickness = 3 $\times$ IDT period). A Si waveguide of width 450 nm and thickness 220 nm was defined underneath the PZT layer, 3 $\mu$m away from the IDT. 

\begin{figure}[ht]
\centering
\includegraphics[width=0.5\linewidth]{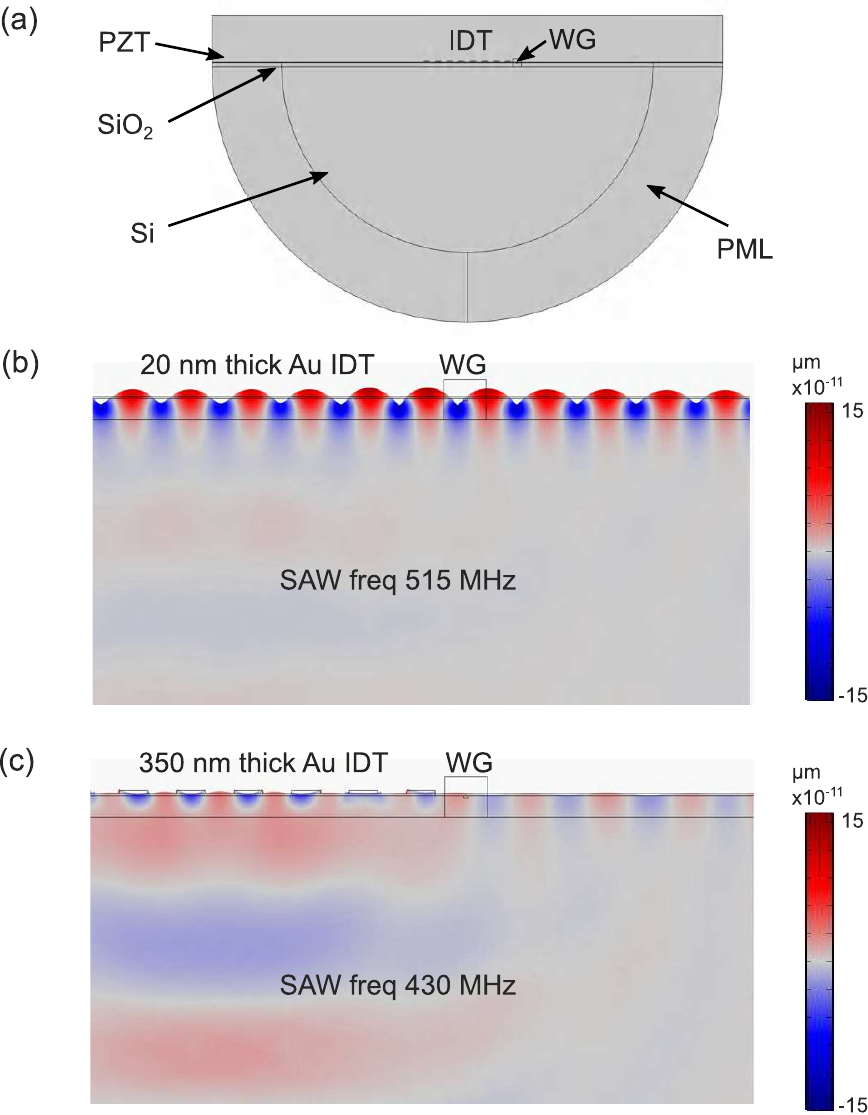}
\caption{(a) 2D simulation setup showing and IDT with 4 finger-pairs defined on a 200 nm thick PZT layer on SOI. A WG is created at 3 $\mu$m spacing from the IDT, underneath the PZT layer. (b,c) Vertical displacement for the fundamental SAW mode of a 20 nm and 350 nm thick Au IDT actuated with 1V amplitude at respectively 515 MHz and 430 MHz.}
\label{fig:SAWthicknessEffectPML}
\end{figure}

In figure \ref{fig:SAWthicknessEffectPML}(b) and (c), we show the fundamental SAW mode launched by a 20 nm thick and 350 nm thick Au IDT (period 12 $\mu$m) respectively. We can see that for the thin IDT, the SAW transduction is stronger and the mode is mostly confined close to the surface. However, with the added mass-loading and grating reflection from a thicker IDT, as shown in \ref{fig:SAWthicknessEffectPML}(c), the SAW resonance frequency is decreased and now more acoustic energy is diffracted into the substrate. 

\begin{figure}[ht]
\centering
\includegraphics[width=0.5\linewidth]{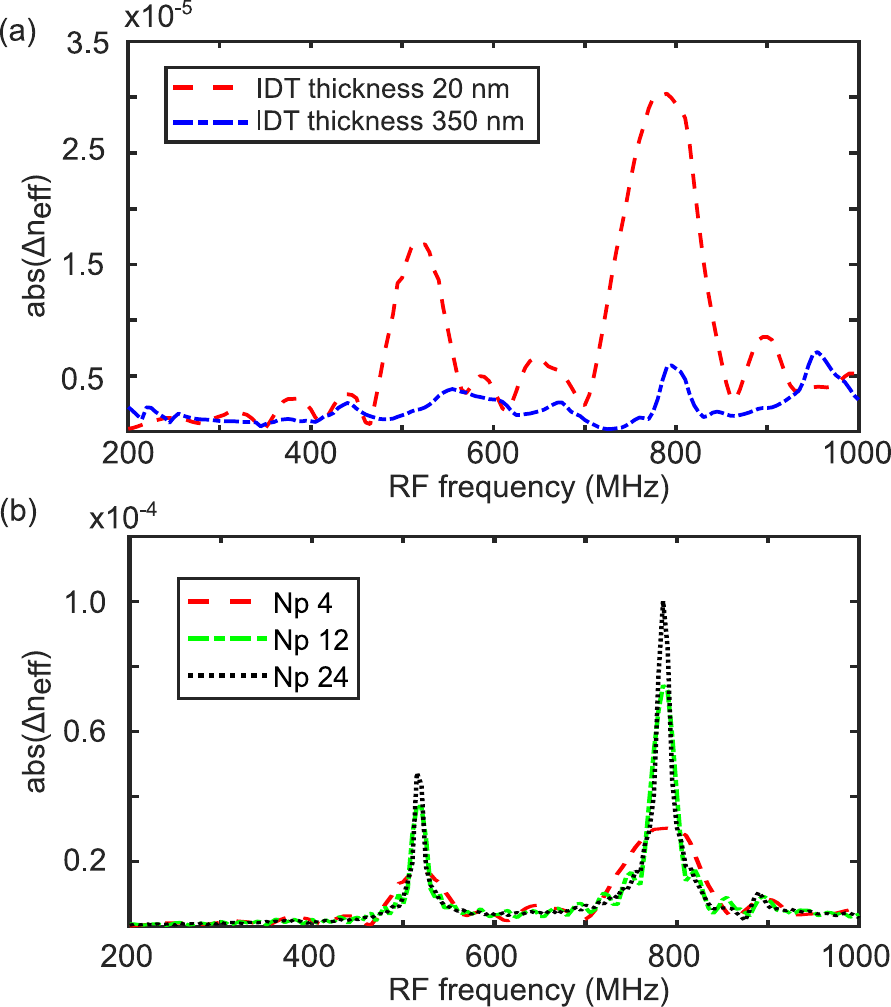}
\caption{$\Delta n_{eff}$ calculated for 12 $\mu$m period Au IDT actuated with a 1V RF signal showing (a) effect of IDT thickness at N$_p$ = 4, (b) effect of N$_p$ at Au IDT thickness 20 nm.}
\label{fig:absdelneff_AuIDT}
\end{figure}

From the simulated strain-fields shown in figure \ref{fig:SAWthicknessEffectPML} and the optical field (TE mode) obtained from a mode analysis, we calculated the acousto-optic overlap integral and extracted the change in effective refractive index $\Delta n_{eff}$ as shown in figure \ref{fig:absdelneff_AuIDT}. This $\Delta n_{eff}$  originates from the photoelastic effect in the waveguide \cite{LiNbO3micro2opto_Loncar2019}.  In our case, since the light propagates along the Si [110] direction, we applied a rotational operator on the photoelastic tensor of Si taken from \cite{Siphotoelastic_biegelsen1975}. Thus we obtained the following coefficients that we used for calculating the overlap integral:
 $p_{11}$= -0.090, $ p_{12}$ = 0.013, $p_{21} $ = $p_{12}$ and $p_{22}$ = $p_{11}$.
 
 In figure \ref{fig:absdelneff_AuIDT}(a), we show the effect of the electrode thickness on $\Delta n_{eff}$. As expected, the index modulation decreases dramatically when the Au IDT thickness is increased from 20nm to 350 nm IDT. In figure \ref{fig:absdelneff_AuIDT}(b), we show the effect of the number of finger-pairs (N$_p$) on $\Delta n_{eff}$ . As expected, the resonance bandwidth decreases and the amplitude increases when N$_p$ is increased \cite{grating_Bandwidth_Erdogan1997}.  

\subsection{Acousto-optic interaction}   \label{appen:ao}

\begin{figure}[ht]
\centering
\includegraphics[width=0.6\linewidth]{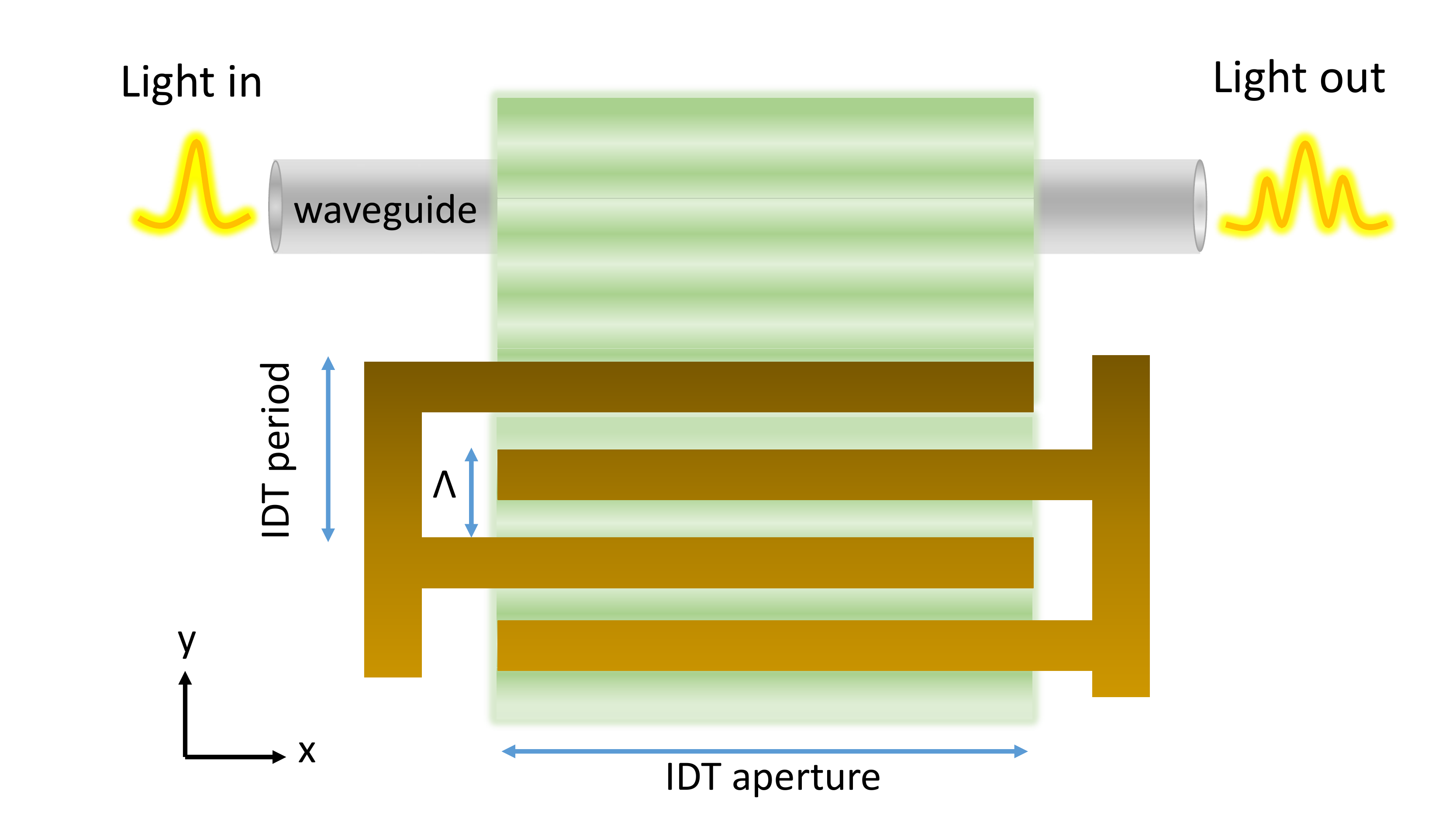}
\caption{The IDT actuated acoustic beam ($\Omega$) diffracts the input waveguide mode ($\omega_0$) into two sidebands ($\omega_0 \pm \Omega$). The wavelength of the fundamental SAW mode $\Lambda$= IDT period/2, for the PZT with the periodic (in-plane) domain orientation.}
\label{fig:AOMdiffraction}
\end{figure}
The electric field of a waveguide mode travelling along $x$ is given by: 
\begin{equation} \label{eqn:mode1}
\psi(x,t)= A\Re e^{i(\omega_0 t-k_0 n_{eff} \, x)} 
\end{equation}
Where $A$ is the electric field amplitude, $\omega_0$ is the angular frequency, $k_0 = 2\pi/\lambda_0$ with $\lambda_0$ as the free space wavelength and \textit{n$_{eff}(\omega_0)$} is the effective refractive index of the waveguide mode. The index modulation in the waveguide material due to an acoustic wave travelling along $y$ (as shown in figure \ref{fig:AOMdiffraction}) can be written as:
\[ n(y,t) = n^0 + \Delta n\, \textrm{sin}\big( \Omega (t-y/v_{ac}) \big) \]
Where $n^0$ is the material index without any modulation, $\Delta n$ is the index change due to the modulated strain from the elastic wave (photoelastic effect). $\Omega$ and $v_{ac}$ are the angular frequency and the phase velocity of the acoustic wave respectively.  These material indices determine the effective index of the waveguide mode which can be approximated as,

\begin{equation} \label{eqn:neff_expression}
 n_{eff}(t) = n_{eff}^0 + \Delta n_{eff}\, \textrm{sin}( \Omega \,t ) 
\end{equation}
where $n_{eff}^0$ is the effective index without any perturbation and $\Delta n_{eff}$ is the amplitude of the effective index change due to the acousto-optic interaction. $\Delta n_{eff}$ depends on the overlap integral between the strain-field and the optical-field \cite{LiNbO3micro2opto_Loncar2019}. From the photoelastic effect, $\Delta n_{eff}$ can be calculated as:
\begin{equation} \label{eqn:photoelastic}
    \Delta n_{eff} = \frac{n_{eff}^3}{2}  \frac{\int_D E^*\boldsymbol{pS} \,E \,dr}{\int_D E^*E \,dr}
\end{equation}
Here $D$ is the 2D cross-section of the waveguide, $E$ is the electric field of the waveguide mode, $\boldsymbol{p}$ is the photoelastic tensor of the waveguide medium and $\boldsymbol{S}$ is the strain field induced by the acoustic wave.

We assume the modulator length is small, so the spatial variation in $n_{eff}$ along the modulator (x-direction) is negligible as $\Omega \, \Delta t = \Omega \,L/v_p  << 1$, where $v_p$ is the light phase velocity.   
Thus, considering no attenuation, the waveguide mode propagating through the modulator can be written as:
\[
\psi(x,t)= A\Re e^{i(\omega_0 t - k_0 \, n_{eff}^0 \, x)} e^{-i k_0\,x\Delta n_{eff}\, \textrm{sin}( \Omega \,t)} \]
\begin{equation} \label{eqn:mode2}
\psi(x,t)= A\Re e^{i(\omega_0 t- k_0 \, n_{eff}^0 \, x)} e^{i\alpha(x) \, \textrm{sin}( \Omega \,t)} 
\end{equation}
Where $\alpha(x) = - k_0 \Delta n_{eff} x$ is the modulated phase amplitude. 
The sinusoidal phase term here can be expanded in terms of the Bessel functions ($J_N$)  
$\implies$  $e^{i\alpha \, \textrm{sin}( \Omega \,t)} =  \sum_{-\infty}^\infty J_N(\alpha)e^{i N\Omega t} $, where $N$ is an integer. Hence, we get the following expression:   

\begin{equation} \label{eqn:modeBessel}
\begin{split}
\psi(x,t)  = A\Re \, e^{i(\omega_0 t - k_0\,  n_{eff}^0 \, x)} \Big[J_0(\alpha(x)) + \sum_{N=1}^\infty J_N(\alpha(x)) \, \big[e^{i N\Omega t} + (-1)^N e^{-i N\Omega t} \big] \Big] 
\end{split}
\end{equation} 
If the modulator length and $\Delta n_{eff}$ are assumed to be small, the phase modulation $\alpha(x) << 1$. Hence, the Bessel function can be simplified as, $J_N(\alpha(x)) \approx \alpha(x)^N/(2^N N!)$. 
Therefore, neglecting the higher order terms in the above equation, 
\begin{equation}  \label{eqn:modeBesselExpanded} 
\begin{split}
\psi(x,t)
 =  A\Re \, e^{-i k_0 \, n_{eff}^0 \, x} \Big[e^{i\omega_0 t} \\
 + \frac{\alpha(x)}{2}
 \big[ \, e^{ i((\omega_0 + \Omega)t  }- e^{ i((\omega_0 - \Omega)t}  \big] \Big]
\end{split}
\end{equation} 
We see that as the carrier mode propagates along the modulator, it is diffracted into Stokes and Antistokes sidebands with frequencies $\omega_0\pm \Omega$. 
The phase obtained by the Antistokes wave generated at $x=0$ accumulated over the modulation length $L$ is then: 
\begin{equation} \label{eqn:phase1}
\begin{split}
\phi_A &= (\omega_0+ \Omega) t_1 \\
       &= (\omega_0+ \Omega) \frac{L}{v_p(\omega_0 + \Omega)} \\
       &= (\omega_0+ \Omega) \frac{n_{eff}(\omega_0 + \Omega) L}{c} \\
\end{split}
\end{equation}

The phase of an Antistokes wave scattered at the end of the modulator on the other hand equals the phase obtained by the carrier wave over the modulation length $L$: 
\begin{equation} \label{eqn:phase2}
\begin{split}
\phi_c &= \omega_0 t_2 \\
       &= \omega_0 \frac{L}{v_p(\omega_0)} \\
       &= \omega_0 \frac{n_{eff}(\omega_0) L}{c} \\
\end{split}
\end{equation}

Hence, the phase difference built-up between these Antistokes sidebands scattered at the beginning ($x=0$) and the end of the modulator ($x=L$):

\begin{equation} \label{eqn:phasediff1}
\begin{split}
\Delta \phi_L &= \phi_A - \phi_c  \\
              &= \frac{L}{c} \Big[ (\omega_0 + \Omega)\, n_{eff}(\omega_0 + \Omega) - \omega_0 \, n_{eff}(\omega_0) \Big] \\
              &= \frac{L\,\omega_0}{c} \Big[ n_{eff}(\omega_0 + \Omega) -  n_{eff}(\omega_0) + \frac{\Omega}{\omega_0} \, n_{eff}(\omega_0 + \Omega) \Big]
\end{split}
\end{equation}
Since $\Omega << \omega_0$,  
$\implies \Omega/\omega_0 \rightarrow 0$. \newline
Also, $n_{eff}(\omega_0 + \Omega) -  n_{eff}(\omega_0) \rightarrow \Omega \frac{\partial n_{eff}}{\partial \omega}\bigg|_{\omega=\omega_0}$

Therefore,
\begin{equation} \label{eqn:phasediff2}
\begin{split}
\Delta \phi_L &= L\,k_0 \,\Omega \frac{\partial n_{eff}}{\partial \omega}\bigg|_{\omega=\omega_0}\\
              &= L\,\Omega \, \frac{\partial k}{\partial \omega}\bigg|_{\omega=\omega_0}\\
              &= L\,\Omega/v_g(\omega_0) \\
\end{split}
\end{equation}

As long as $L$ is small enough, this phase difference is negligible ($\Delta \phi_L << \pi$), and the diffracted signals will add up constructively producing a carrier peak and two (modulated) sideband peaks in the output spectra. This is reminiscent of Raman-Nath diffraction of a light beam in free space due to an elastic beam propagation \cite{royerbook2}. 

For our modulators $\Delta \phi_L$ was indeed negligible. For instance, with an IDT of aperture length ($L$) 180 $\mu$m and acoustic wave frequency ($\Omega$) 576 MHz, $\Delta \phi_L$ was estimated to be $\approx$ 0.008 $<< \pi$. We used $v_g(\omega_0) \approx$ 1.3 x 10$^7$ m/s calculated from the single mode (TE00) dispersion relation which was obtained with Comsol mode solver. 

It is noteworthy that a high phase-modulation ($\alpha(L)$) can be achieved by designing a cascaded modulator system. For example, a serpentine waveguide structure can be designed next to the IDT to allow multiple (smaller) interaction-lengths \cite{SAW_AlNSOI_Kittlaus2021, sawthermo1-Avi2019}. Hence, the net phase-modulation accumulated over the total interaction-lengths can be very high, while the phase-mismatch ($\Delta \phi_L$) in the individual sections remains negligible. 



\section{Acknowledgments}
This work was supported by the EU commission through Grant agreement No. 732894  (FET proactive HOT) and through the UGent grant BOFGOA2020000103. Gilles F. Feutmba acknowledges support and funding as an SB-PhD Fellow of the research foundation–Flanders (FWO, grant number 1S68218N).

\section{Disclosures}
The authors declare no conflict of interest

\bibliographystyle{unsrt}
\bibliography{Main}

\begin{thebibliography}{10}

\bibitem{Marpaung2013}
David Marpaung, Chris Roeloffzen, René Heideman, Arne Leinse, Salvador Sales,
  and José Capmany.
\newblock Integrated microwave photonics.
\newblock {\em Laser \& Photonics Reviews}, 7(4):506--538, 2013.

\bibitem{marpaung2015low}
David Marpaung, Blair Morrison, Mattia Pagani, Ravi Pant, Duk-Yong Choi, Barry
  Luther-Davies, Steve~J Madden, and Benjamin~J Eggleton.
\newblock Low-power, chip-based stimulated brillouin scattering microwave
  photonic filter with ultrahigh selectivity.
\newblock {\em Optica}, 2(2):76--83, 2015.

\bibitem{kang2011reconfigurable}
Myeong~Soo Kang, A~Butsch, and P~St~J Russell.
\newblock Reconfigurable light-driven opto-acoustic isolators in photonic
  crystal fibre.
\newblock {\em Nature Photonics}, 5(9):549--553, 2011.

\bibitem{isolator_yu2009}
Zongfu Yu and Shanhui Fan.
\newblock Complete optical isolation created by indirect interband photonic
  transitions.
\newblock {\em Nature photonics}, 3(2):91--94, 2009.

\bibitem{isolator_huang2016}
Duanni Huang, Paolo Pintus, Chong Zhang, Yuya Shoji, Tetsuya Mizumoto, and
  John~E Bowers.
\newblock Electrically driven and thermally tunable integrated optical
  isolators for silicon photonics.
\newblock {\em IEEE Journal of Selected Topics in Quantum Electronics},
  22(6):271--278, 2016.

\bibitem{nonrecipmodulator_sohn2019}
Donggyu~B Sohn and Gaurav Bahl.
\newblock Direction reconfigurable nonreciprocal acousto-optic modulator on
  chip.
\newblock {\em APL Photonics}, 4(12):126103, 2019.

\bibitem{modulator_fan2016integrated}
Linran Fan, Chang-Ling Zou, Menno Poot, Risheng Cheng, Xiang Guo, Xu~Han, and
  Hong~X Tang.
\newblock Integrated optomechanical single-photon frequency shifter.
\newblock {\em Nature Photonics}, 10(12):766--770, 2016.

\bibitem{modulator_tadesse2014sub}
Semere~Ayalew Tadesse and Mo~Li.
\newblock Sub-optical wavelength acoustic wave modulation of integrated
  photonic resonators at microwave frequencies.
\newblock {\em Nature communications}, 5(1):1--7, 2014.

\bibitem{modulator_tadesse2015acousto}
Semere~A Tadesse, Huan Li, Qiyu Liu, and Mo~Li.
\newblock Acousto-optic modulation of a photonic crystal nanocavity with lamb
  waves in microwave k band.
\newblock {\em Applied Physics Letters}, 107(20):201113, 2015.

\bibitem{modulator_balram2017acousto}
Krishna~C Balram, Marcelo~I Davan{\c{c}}o, B~Robert Ilic, Ji-Hoon Kyhm,
  Jin~Dong Song, and Kartik Srinivasan.
\newblock Acousto-optic modulation and optoacoustic gating in
  piezo-optomechanical circuits.
\newblock {\em Physical review applied}, 7(2):024008, 2017.

\bibitem{modulator_de2006compact}
MM~de~Lima~Jr, M~Beck, R~Hey, and PV~Santos.
\newblock Compact mach-zehnder acousto-optic modulator.
\newblock {\em Applied physics letters}, 89(12):121104, 2006.

\bibitem{LiNbO3_Hassanien2021}
Ahmed~E. Hassanien, Steffen Link, Yansong Yang, Edmond Chow, Lynford~L.
  Goddard, and Songbin Gong.
\newblock Efficient and wideband acousto-optic modulation on thin-film lithium
  niobate for microwave-to-photonic conversion.
\newblock {\em Photon. Res.}, 9(7):1182--1190, Jul 2021.

\bibitem{LiNbO3_Sun2021}
Zejie Yu and Xiankai Sun.
\newblock Gigahertz acousto-optic modulation and frequency shifting on etchless
  lithium niobate integrated platform.
\newblock {\em ACS Photonics}, 8(3):798--803, 2021.

\bibitem{modeshift_kuhn1971}
L~Kuhn, PF~Heidrich, and EG~Lean.
\newblock Optical guided wave mode conversion by an acoustic surface wave.
\newblock {\em Applied Physics Letters}, 19(10):428--430, 1971.

\bibitem{modeshift_sasaki1974}
H~Sasaki, J~Kushibiki, and N~Chubachi.
\newblock Efficient acousto-optic te-tm mode conversion in zno films.
\newblock {\em Applied Physics Letters}, 25(9):476--477, 1974.

\bibitem{modeshift_ohmachi1977}
YOSHIRO Ohmachi and JUICHI Noda.
\newblock Linbo 3 te-tm mode converter using collinear acoustooptic
  interaction.
\newblock {\em IEEE Journal of Quantum Electronics}, 13(2):43--46, 1977.

\bibitem{phonon-photon_Safavi19}
Amir~H. Safavi-Naeini, Dries~Van Thourhout, Roel Baets, and Rapha\"{e}l~Van
  Laer.
\newblock Controlling phonons and photons at the wavelength scale: integrated
  photonics meets integrated phononics.
\newblock {\em Optica}, 6(2):213--232, Feb 2019.

\bibitem{timreversal_Bahl2018}
Donggyu~B. Sohn, Seunghwi Kim, and Gaurav Bahl.
\newblock {Time-reversal symmetry breaking with acoustic pumping of
  nanophotonic circuits}.
\newblock {\em Nature Photonics}, 12(2):91--97, feb 2018.

\bibitem{nonreciprocalBrillouin_Rakich2018}
Eric~A. Kittlaus, Nils~T. Otterstrom, Prashanta Kharel, Shai Gertler, and
  Peter~T. Rakich.
\newblock {Non-reciprocal interband Brillouin modulation}.
\newblock {\em Nature Photonics}, 12(10):613--619, oct 2018.

\bibitem{SAW_AlNSOI_Kittlaus2021}
Eric~A. Kittlaus, William~M. Jones, Peter~T. Rakich, Nils~T. Otterstrom,
  Richard~E. Muller, and Mina Rais-Zadeh.
\newblock Electrically driven acousto-optics and broadband non-reciprocity in
  silicon photonics.
\newblock {\em Nature Photonics}, 15:43--52, 1 2021.

\bibitem{freqcomb_Jessen92}
Poul Jessen and Martin Kristensen.
\newblock Generation of a frequency comb with a double acousto-optic modulator
  ring.
\newblock {\em Appl. Opt.}, 31(24):4911--4913, Aug 1992.

\bibitem{freqcomb_Carlos19}
Vicente Duran, Hugues~Guillet De~Chatellus, Come Schnebelin, Kanagaraj
  Nithyanandan, Leo Djevarhidjian, Juan Clement, and Carlos~R. Fernandez-Pousa.
\newblock Optical frequency combs generated by acousto-optic frequency-shifting
  loops.
\newblock {\em IEEE Photonics Technology Letters}, 31(23):1878--1881, 2019.

\bibitem{freqcomb_Loncar20}
Linbo Shao, Neil Sinclair, James Leatham, Yaowen Hu, Mengjie Yu, Terry Turpin,
  Devon Crowe, and Marko Lon\v{c}ar.
\newblock Integrated microwave acousto-optic frequency shifter on thin-film
  lithium niobate.
\newblock {\em Opt. Express}, 28(16):23728--23738, Aug 2020.

\bibitem{RevModPhys_kippenberg2014}
Markus Aspelmeyer, Tobias~J. Kippenberg, and Florian Marquardt.
\newblock Cavity optomechanics.
\newblock {\em Rev. Mod. Phys.}, 86:1391--1452, Dec 2014.

\bibitem{andrews2014bidirectional}
Reed~W Andrews, Robert~W Peterson, Tom~P Purdy, Katarina Cicak, Raymond~W
  Simmonds, Cindy~A Regal, and Konrad~W Lehnert.
\newblock Bidirectional and efficient conversion between microwave and optical
  light.
\newblock {\em Nature physics}, 10(4):321--326, 2014.

\bibitem{verhagen2012quantum}
Ewold Verhagen, Samuel Del{\'e}glise, Stefan Weis, Albert Schliesser, and
  Tobias~J Kippenberg.
\newblock Quantum-coherent coupling of a mechanical oscillator to an optical
  cavity mode.
\newblock {\em Nature}, 482(7383):63--67, 2012.

\bibitem{QAO_Nysten_2017}
Emeline D~S Nysten, Yong~Heng Huo, Hailong Yu, Guo~Feng Song, Armando Rastelli,
  and Hubert~J Krenner.
\newblock Multi-harmonic quantum dot optomechanics in fused
  {LiNbO}3{\textendash}(al){GaAs} hybrids.
\newblock {\em Journal of Physics D: Applied Physics}, 50(43):43LT01, sep 2017.

\bibitem{GaAs_kapfinger2015dynamic}
Stephan Kapfinger, Thorsten Reichert, Stefan Lichtmannecker, Kai M{\"u}ller,
  Jonathan~J Finley, Achim Wixforth, Michael Kaniber, and Hubert~J Krenner.
\newblock Dynamic acousto-optic control of a strongly coupled photonic
  molecule.
\newblock {\em Nature communications}, 6(1):1--6, 2015.

\bibitem{LiNbO3_Cai2019}
Lutong Cai, Ashraf Mahmoud, and Gianluca Piazza.
\newblock Low-loss waveguides on y-cut thin film lithium niobate: towards
  acousto-optic applications.
\newblock {\em Opt. Express}, 27(7):9794--9802, Apr 2019.

\bibitem{LiNbO3micro2opto_Loncar2019}
Linbo Shao, Mengjie Yu, Smarak Maity, Neil Sinclair, Lu~Zheng, Cleaven Chia,
  Amirhassan Shams-Ansari, Cheng Wang, Mian Zhang, Keji Lai, and Marko
  Lon\v{c}ar.
\newblock Microwave-to-optical conversion using lithium niobate thin-film
  acoustic resonators.
\newblock {\em Optica}, 6(12):1498--1505, Dec 2019.

\bibitem{LiNbO3_Safavi2020}
Christopher~J. Sarabalis, Timothy~P. McKenna, Rishi~N. Patel, Raphaël
  Van~Laer, and Amir~H. Safavi-Naeini.
\newblock Acousto-optic modulation in lithium niobate on sapphire.
\newblock {\em APL Photonics}, 5(8):086104, 2020.

\bibitem{compactmziGaAs_delima2006}
M.~M. de~Lima, M.~Beck, R.~Hey, and P.~V. Santos.
\newblock Compact mach-zehnder acousto-optic modulator.
\newblock {\em Applied Physics Letters}, 89(12):121104, 2006.

\bibitem{aom_delima2007}
M.~Beck, M.~M. de~Lima, E.~Wiebicke, W.~Seidel, R.~Hey, and P.~V. Santos.
\newblock Acousto-optical multiple interference switches.
\newblock {\em Applied Physics Letters}, 91(6):061118, 2007.

\bibitem{GaAs_Kartik2017}
Krishna~C. Balram, Marcelo~I. Davan\ifmmode~\mbox{\c{c}}\else \c{c}\fi{}o,
  B.~Robert Ilic, Ji-Hoon Kyhm, Jin~Dong Song, and Kartik Srinivasan.
\newblock Acousto-optic modulation and optoacoustic gating in
  piezo-optomechanical circuits.
\newblock {\em Phys. Rev. Applied}, 7:024008, Feb 2017.

\bibitem{InP_Sun95}
B.~Sun, A.~Kar-Roy, and C.~S. Tsai.
\newblock Guided-wave acousto-optic bragg diffractions in inp/ingaasp/inp
  waveguide.
\newblock In {\em Integrated Photonics Research}, page IThG15. Optical Society
  of America, 1995.

\bibitem{InP1_Makles2015}
K.~Makles, T.~Antoni, A.~G. Kuhn, S.~Del\'{e}glise, T.~Briant, P.-F. Cohadon,
  R.~Braive, G.~Beaudoin, L.~Pinard, C.~Michel, V.~Dolique, R.~Flaminio,
  G.~Cagnoli, I.~Robert-Philip, and A.~Heidmann.
\newblock 2d photonic-crystal optomechanical nanoresonator.
\newblock {\em Opt. Lett.}, 40(2):174--177, Jan 2015.

\bibitem{Bahram_Siphotonic2006}
Bahram Jalali and Sasan Fathpour.
\newblock Silicon photonics.
\newblock {\em Journal of Lightwave Technology}, 24(12):4600--4615, 2006.

\bibitem{SiCMOS_Bogaerts2005}
W.~Bogaerts, R.~Baets, P.~Dumon, V.~Wiaux, S.~Beckx, D.~Taillaert,
  B.~Luyssaert, J.~Van~Campenhout, P.~Bienstman, and D.~Van~Thourhout.
\newblock Nanophotonic waveguides in silicon-on-insulator fabricated with cmos
  technology.
\newblock {\em Journal of Lightwave Technology}, 23(1):401--412, 2005.

\bibitem{sawthermo1-Avi2019}
Dvir Munk, Moshe Katzman, Mirit Hen, Maayan Priel, Moshe Feldberg, Tali
  Sharabani, Shahar Levy, Arik Bergman, and Avi Zadok.
\newblock {Surface acoustic wave photonic devices in silicon on insulator}.
\newblock {\em Nature Communications}, 10(1):4214, dec 2019.

\bibitem{sawthermo2-Avi2021}
Moshe Katzman, Dvir Munk, Maayan Priel, Etai Grunwald, Mirit Hen, Naor Inbar,
  Moshe Feldberg, Tali Sharabani, Roy Zektzer, Gil Bashan, Menachem Vofsi,
  Uriel Levy, and Avi Zadok.
\newblock Surface acoustic microwave photonic filters in standard
  silicon-on-insulator.
\newblock {\em Optica}, 8(5):697--707, May 2021.

\bibitem{AlNforSOI_Tang2012}
Chi Xiong, Wolfram H~P Pernice, Xiankai Sun, Carsten Schuck, King~Y Fong, and
  Hong~X Tang.
\newblock Aluminum nitride as a new material for chip-scale optomechanics and
  nonlinear optics.
\newblock {\em New Journal of Physics}, 14(9):095014, sep 2012.

\bibitem{auld1}
Bertram~Alexander Auld.
\newblock {\em Acoustic fields and waves in solids}.
\newblock John Wiley \& Sons, 1973.

\bibitem{piezoreview_2019}
Mohsen Safaei, Henry~A Sodano, and Steven~R Anton.
\newblock A review of energy harvesting using piezoelectric materials:
  state-of-the-art a decade later (2008{\textendash}2018).
\newblock {\em Smart Materials and Structures}, 28(11):113001, oct 2019.

\bibitem{piezoceramic_1958}
HANS JAFFE.
\newblock Piezoelectric ceramics.
\newblock {\em Journal of the American Ceramic Society}, 41(11):494--498, 1958.

\bibitem{lossyPZT2017}
N.~Izyumskaya, Y.-I. Alivov, S.-J. Cho, H.~Morkoç, H.~Lee, and Y.-S. Kang.
\newblock Processing, structure, properties, and applications of pzt thin
  films.
\newblock {\em Critical Reviews in Solid State and Materials Sciences},
  32(3-4):111--202, 2007.

\bibitem{hosseini2015}
Naser Hosseini, Ronald Dekker, Marcel Hoekman, Matthijn Dekkers, Jan Bos, Arne
  Leinse, and Rene Heideman.
\newblock Stress-optic modulator in triplex platform using a piezoelectric lead
  zirconate titanate (pzt) thin film.
\newblock {\em Opt. Express}, 23(11):14018--14026, Jun 2015.

\bibitem{pzt_George2015}
J.~P. George, P.~F. Smet, J.~Botterman, V.~Bliznuk, W.~Woestenborghs, D.~{Van
  Thourhout}, K.~Neyts, and J.~Beeckman.
\newblock {Lanthanide-Assisted Deposition of Strongly Electro-optic PZT Thin
  Films on Silicon: Toward Integrated Active Nanophotonic Devices}.
\newblock {\em ACS Applied Materials \& Interfaces}, 7(24):13350--13359, jun
  2015.

\bibitem{Alexander2018}
Koen Alexander, John~P. George, Jochem Verbist, Kristiaan Neyts, Bart Kuyken,
  Dries {Van Thourhout}, and Jeroen Beeckman.
\newblock {Nanophotonic Pockels modulators on a silicon nitride platform}.
\newblock {\em Nature Communications}, 9(1):4--9, 2018.

\bibitem{Gilles_EO2020}
Gilles~Freddy Feutmba, Tessa~Van de~Veire, Irfan Ansari, John~P. George,
  Dries~Van Thourhout, and Jeroen Beeckman.
\newblock A strong pockels pzt/si modulator for efficient electro-optic tuning.
\newblock In {\em OSA Advanced Photonics Congress (AP) 2020 (IPR, NP, NOMA,
  Networks, PVLED, PSC, SPPCom, SOF)}, page ITu1A.6. Optical Society of
  America, 2020.

\bibitem{Gilles_nonlinearPZT2021}
Gilles~F Feutmba, Artur Hermans, John~P George, Hannes Rijckaert, Irfan Ansari,
  Dries Van~Thourhout, and Jeroen Beeckman.
\newblock Reversible and tunable second-order nonlinear optical susceptibility
  in pzt thin films for integrated optics.
\newblock {\em Advanced Optical Materials}, 9(16):2100149, 2021.

\bibitem{sawpzt_cleo2020}
Irfan Ansari, Tessa~Van de~Veire, John~P. George, Gilles.F. Feutmba, Jeroen
  Beeckman, and Dries~Van Thourhout.
\newblock Si-photonic integrated pzt thin film for acousto-optic modulation.
\newblock In {\em Conference on Lasers and Electro-Optics}, page JTh2B.24.
  Optical Society of America, 2020.

\bibitem{pztAOsim_Klaus2019}
Peter J.~M. van~der Slot, Marco A.~G. Porcel, and Klaus-J. Boller.
\newblock Surface acoustic waves for acousto-optic modulation in buried silicon
  nitride waveguides.
\newblock {\em Opt. Express}, 27(2):1433--1452, Jan 2019.

\bibitem{GHzPZT_Irfan2021}
Irfan Ansari, Dries~Van Thourhout, John~P. George, Gilles~F. Feutmba, and
  Jeroen Beeckman.
\newblock Acousto-optic modulation in a si-waveguide.
\newblock In {\em 2021 IEEE 17th International Conference on Group IV Photonics
  (GFP)}, pages 1--2, 2021.

\bibitem{delima_sawmod2013}
A.~Crespo-Poveda, R.~Hey, K.~Biermann, A.~Tahraoui, P.~V. Santos, B.~Gargallo,
  P.~Mu\ {n}oz, A.~Cantarero, and M.~M. de~Lima.
\newblock Synchronized photonic modulators driven by surface acoustic waves.
\newblock {\em Opt. Express}, 21(18):21669--21676, Sep 2013.

\bibitem{rayleigh}
Lord Rayleigh.
\newblock On waves propagated along the plane surface of an elastic solid.
\newblock {\em Proceedings of the London Mathematical Society}, s1-17(1):4--11,
  1885.

\bibitem{sawidt1977}
R.F. Milsom, N.H.C. Reilly, and M.~Redwood.
\newblock Analysis of generation and detection of surface and bulk acoustic
  waves by interdigital transducers.
\newblock {\em IEEE Transactions on Sonics and Ultrasonics}, 24(3):147--166,
  1977.

\bibitem{sawmat}
A.J. Slobodnik.
\newblock Surface acoustic waves and saw materials.
\newblock {\em Proceedings of the IEEE}, 64(5):581--595, 1976.

\bibitem{sawthermo_0}
C.~Giannetti, B.~Revaz, F.~Banfi, M.~Montagnese, G.~Ferrini, F.~Cilento,
  S.~Maccalli, P.~Vavassori, G.~Oliviero, E.~Bontempi, L.~E. Depero,
  V.~Metlushko, and F.~Parmigiani.
\newblock Thermomechanical behavior of surface acoustic waves in ordered arrays
  of nanodisks studied by near-infrared pump-probe diffraction experiments.
\newblock {\em Phys. Rev. B}, 76:125413, Sep 2007.

\bibitem{sawthermo_1}
Damiano Nardi, Marco Travagliati, Mark~E. Siemens, Qing Li, Margaret~M.
  Murnane, Henry~C. Kapteyn, Gabriele Ferrini, Fulvio Parmigiani, and Francesco
  Banfi.
\newblock Probing thermomechanics at the nanoscale: Impulsively excited
  pseudosurface acoustic waves in hypersonic phononic crystals.
\newblock {\em Nano Letters}, 11(10):4126--4133, 2011.
\newblock PMID: 21910426.

\bibitem{sawthermo_2}
Martin Schubert, Martin Grossmann, Oliver Ristow, Mike Hettich, Axel
  Bruchhausen, Elaine C.~S. Barretto, Elke Scheer, Vitalyi Gusev, and Thomas
  Dekorsy.
\newblock Spatial-temporally resolved high-frequency surface acoustic waves on
  silicon investigated by femtosecond spectroscopy.
\newblock {\em Applied Physics Letters}, 101(1):013108, 2012.

\bibitem{sawdesign}
Colin Campbell.
\newblock {\em Surface Acoustic Wave Devices for Mobile and Wireless
  Communications, Four-Volume Set}.
\newblock Academic press, 1998.

\bibitem{royerbook2}
Daniel Royer and Eugene Dieulesaint.
\newblock {\em {Elastic waves in solids II: generation, acousto-optic
  interaction, applications}}.
\newblock Springer Science \& Business Media, 1999.

\bibitem{30GHz_leakySAWlinbo3}
Jiangpo Zheng, Jian Zhou, Pei Zeng, Yi~Liu, Yiping Shen, Wenze Yao, Zhe Chen,
  Jianhui Wu, Shuo Xiong, Yiqin Chen, Xianglong Shi, Jie Liu, Yongqing Fu, and
  Huigao Duan.
\newblock 30 ghz surface acoustic wave transducers with extremely high mass
  sensitivity.
\newblock {\em Applied Physics Letters}, 116(12):123502, 2020.

\bibitem{leakySAWquartz}
Masahiro Oshio, Shigeo Kanna, and Keigo Iizawa.
\newblock Effect of substrate thickness on quasi-longitudinal leaky saw
  propagation on quartz.
\newblock In {\em IEEE Ultrasonics Symposium}, volume~4, pages 1880--1883.
  IEEE, 2005.

\bibitem{PZTdomain_Kratzer2018}
Markus Kratzer, Michael Lasnik, S{\"{o}}ren R{\"{o}}hrig, Christian Teichert,
  and Marco Deluca.
\newblock Reconstruction of the domain orientation distribution function of
  polycrystalline pzt ceramics using vector piezoresponse force microscopy.
\newblock {\em Scientific Reports 2017 8:1}, 8(1):1--11, jan 2018.

\bibitem{piezoIEEEstd}
{IEEE Standard on Piezoelectricity}.
\newblock {\em ANSI/IEEE Std 176-1987}, page 0\_1, 1988.

\bibitem{pzt5a}
Zubair Butt, Riffat~Asim Pasha, Faisal Qayyum, Zeeshan Anjum, Nasir Ahmad, and
  Hassan Elahi.
\newblock Generation of electrical energy using lead zirconate titanate
  (pzt-5a) piezoelectric material: Analytical, numerical and experimental
  verifications.
\newblock {\em Journal of Mechanical Science and Technology}, 30(8):3553--3558,
  2016.

\bibitem{johnPZT}
J.~P. George, P.~F. Smet, J.~Botterman, V.~Bliznuk, W.~Woestenborghs,
  D.~Van~Thourhout, K.~Neyts, and J.~Beeckman.
\newblock Lanthanide-assisted deposition of strongly electro-optic pzt thin
  films on silicon: Toward integrated active nanophotonic devices.
\newblock {\em ACS Applied Materials \& Interfaces}, 7(24):13350--13359, 2015.

\bibitem{massloading_chen2020}
Zhe Chen, Jian Zhou, Hao Tang, Yi~Liu, Yiping Shen, Xiaobo Yin, Jiangpo Zheng,
  Hongshuai Zhang, Jianhui Wu, Xianglong Shi, et~al.
\newblock Ultrahigh-frequency surface acoustic wave sensors with giant
  mass-loading effects on electrodes.
\newblock {\em ACS sensors}, 5(6):1657--1664, 2020.

\bibitem{splitfinger-IDT_delima2013}
Antonio Crespo-Poveda, R~Hey, K~Biermann, A~Tahraoui, PV~Santos, B~Gargallo,
  P~Mu{\~n}oz, A~Cantarero, and MM~de~Lima.
\newblock Synchronized photonic modulators driven by surface acoustic waves.
\newblock {\em Optics express}, 21(18):21669--21676, 2013.

\bibitem{uniSAW1}
Maria~K. Ekström, Thomas Aref, Johan Runeson, Johan Björck, Isac Boström,
  and Per Delsing.
\newblock Surface acoustic wave unidirectional transducers for quantum
  applications.
\newblock {\em Applied Physics Letters}, 110(7):073105, 2017.

\bibitem{uniSAW2}
E.~Dumur, K.~J. Satzinger, G.~A. Peairs, Ming-Han Chou, A.~Bienfait, H.-S.
  Chang, C.~R. Conner, J.~Grebel, R.~G. Povey, Y.~P. Zhong, and A.~N. Cleland.
\newblock Unidirectional distributed acoustic reflection transducers for
  quantum applications.
\newblock {\em Applied Physics Letters}, 114(22):223501, 2019.

\bibitem{Siphotoelastic_biegelsen1975}
DK~Biegelsen.
\newblock Frequency dependence of the photoelastic coefficients of silicon.
\newblock {\em Physical Review B}, 12(6):2427, 1975.

\bibitem{grating_Bandwidth_Erdogan1997}
T.~Erdogan.
\newblock Fiber grating spectra.
\newblock {\em Journal of Lightwave Technology}, 15(8):1277--1294, 1997.

\end{thebibliography}

\end{document}